%% file: main.tex
\documentclass[headings=standardclasses]{scrartcl}

\usepackage{geometry}
\usepackage[utf8]{inputenc}
\usepackage[T1]{fontenc}
\usepackage[onehalfspacing]{setspace}
\usepackage{charter}
\usepackage[charter]{mathdesign}
\usepackage{microtype}
\usepackage{amsmath}
\usepackage{amsthm}
\usepackage{tabulary}
\usepackage{longtable}
\usepackage{booktabs} 
\usepackage{diagbox}
\usepackage{ragged2e}
\newcommand{\ra}[1]{\renewcommand{\arraystretch}{#1}}
\usepackage{siunitx}
\usepackage{bm}
\usepackage[pdftex]{graphicx}
\usepackage{rotating}
\usepackage{pdfpages}
\usepackage{pdflscape}
\usepackage{color}
\usepackage{array}
\usepackage[font=bf]{caption}
\usepackage{subcaption}
\usepackage{pgfplots}
\usepackage{pgfplotstable}
\usepackage{float}
\usepackage[title]{appendix}
\pgfplotsset{compat=1.9}
\usepackage{url}
\usepackage[section]{placeins}
\usepackage{natbib}
\usepackage{hyperref}
\hypersetup{
	colorlinks   = true,    
	urlcolor     = blue,    
	linkcolor    = blue,    
	citecolor    = blue      
}

\usepackage[boxed]{algorithm2e}
\SetKwFunction{KwInit}{Initialisation}
\SetKwComment{Comment}{\% }{}

\makeatletter
\renewcommand*\env@matrix[1][c]{\hskip -\arraycolsep
  \let\@ifnextchar\new@ifnextchar
  \array{*\c@MaxMatrixCols #1}}
\makeatother

\usepackage{xcolor}
\usepackage{soul}

\usepackage{verbatim}

\begin{document}
\thispagestyle{empty}
\begin{spacing}{1.2}
\begin{flushleft}
\huge \textbf{Variational Bayesian Inference for Mixed Logit Models with Unobserved Inter- and Intra-Individual Heterogeneity} \\ 
\vspace{\baselineskip}
\normalsize
15 January 2020 \\
\vspace{\baselineskip}
\textsc{Rico Krueger} \\
Transport and Mobility Laboratory \\
Ecole Polytechnique F\'{e}d\'{e}rale de Lausanne, Switzerland \\
rico.krueger@epfl.ch \\
\vspace{\baselineskip}
\textsc{Prateek Bansal} \\
Transport Strategy Centre, Department of Civil and Environmental Engineering\\
Imperial College London, UK \\
prateek.bansal@imperial.ac.uk \\
\vspace{\baselineskip}
\textsc{Michel Bierlaire} \\
Transport and Mobility Laboratory \\
Ecole Polytechnique F\'{e}d\'{e}rale de Lausanne, Switzerland \\
michel.bierlaire@epfl.ch \\
\vspace{\baselineskip}
\textsc{Ricardo A. Daziano} \\
School of Civil and Environmental Engineering \\
Cornell University, United States  \\
daziano@cornell.edu \\
\vspace{\baselineskip}
\textsc{Taha H. Rashidi} (corresponding author) \\
Research Centre for Integrated Transport Innovation, School of Civil and Environmental Engineering
University of New South Wales, Australia\\
rashidi@unsw.edu.au\\
\end{flushleft}
\end{spacing}

\newpage
\thispagestyle{empty}
\section*{Abstract}

Variational Bayes (VB), a method originating from machine learning, enables fast and scalable estimation of complex probabilistic models. Thus far, applications of VB in discrete choice analysis have been limited to mixed logit models with unobserved inter-individual taste heterogeneity. However, such a model formulation may be too restrictive in panel data settings, since tastes may vary both between individuals as well as across choice tasks encountered by the same individual. In this paper, we derive a VB method for posterior inference in mixed logit models with unobserved inter- and intra-individual heterogeneity. In a simulation study, we benchmark the performance of the proposed VB method against maximum simulated likelihood (MSL) and Markov chain Monte Carlo (MCMC) methods in terms of parameter recovery, predictive accuracy and computational efficiency. The simulation study shows that VB can be a fast, scalable and accurate alternative to MSL and MCMC estimation, especially in applications in which fast predictions are paramount. VB is observed to be between 2.8 and 17.7 times faster than the two competing methods, while affording comparable or superior accuracy. Besides, the simulation study demonstrates that a parallelised implementation of the MSL estimator with analytical gradients is a viable alternative to MCMC in terms of both estimation accuracy and computational efficiency, as the MSL estimator is observed to be between 0.9 and 2.1 times faster than MCMC. 
\\
\\
\textit{Keywords:} Variational Bayes; Bayesian inference; mixed logit; inter- and intra-individual heterogeneity.


\newpage
\pagenumbering{arabic}

\section{Introduction}

The representation of taste heterogeneity is a principal concern of discrete choice analysis, as information on the distribution of tastes is critical for demand forecasting, welfare analysis and market segmentation \citep[e.g.][]{allenby1998marketing, ben2019foundations}. From the analyst's perspective, taste variation is often random, as differences in sensitivities cannot be related to observed or observable characteristics of the decision-maker or features of the choice context \citep[see e.g][]{bhat1998accommodating, bhat2000incorporating}. 

Mixed random utility models such as mixed logit \citep{mcfadden2000mixed} provide a powerful framework to account for unobserved taste heterogeneity in discrete choice models. When longitudinal choice data are analysed with the help of mixed random utility models, it is standard practice to assume that tastes vary randomly across decision-makers but not across choice occasions encountered by the same individual \citep{revelt1998mixed}. The implicit assumption underlying this treatment of unobserved heterogeneity is that an individual's tastes are unique and stable \citep{stigler1977gustibus}. Contrasting views of preference formation postulate that preferences are constructed in an ad-hoc manner at the moment of choice \citep{bettman1998constructive} or learnt and discovered through experience \citep{kivetz2008synthesis}. 

From the perspective of discrete choice analysis, these alternative views of preference formation justify accounting for both inter- and intra-individual random heterogeneity \citep[also see][]{hess2015intra}. A straightforward way to accommodate unobserved inter- and intra-individual heterogeneity in mixed random utility models is to augment a mixed logit model with a multivariate normal mixing distribution in a hierarchical fashion such that the case-specific taste parameters are generated as normal perturbations around the individual-specific taste parameters \citep[see][]{becker2018bayesian, bhat2002unified, bhat2006impact, danaf2019online, hess2009allowing, hess2011recovery, hess2015intra, yanez2011treatment}. Besides, \citet{bhat2011simulation} accommodate random parameters with unobserved inter- and inter-individual within the multinomial probit framework and employ the maximum approximate composite marginal likelihood \citep[MACML;][]{bhat2011maximum} approach for model estimation.

Mixed logit models with unobserved inter- and intra-individual heterogeneity can be estimated with the help of maximum simulated likelihood (MSL) estimation methods \citep[in particular][]{hess2009allowing, hess2011recovery}. However, this estimation strategy is computationally expensive, as it involves the simulation of iterated integrals. \citet{becker2018bayesian} propose a Markov chain Monte Carlo (MCMC) method, which builds on the Allenby-Train procedure \citep{train2009discrete} for mixed logit models with only inter-individual heterogeneity. Notwithstanding that MCMC methods constitute a powerful framework for posterior inference in complex probabilistic models \citep[e.g.][]{gelman2013bayesian}, they are subject to several bottlenecks, which inhibit their scalability to large datasets, namely long computation times, serial correlation, high storage costs for the posterior draws and difficulties in assessing convergence \citep[also see][]{bansal2020bayesian, depraetere2017comparison}.

Variational Bayes methods \citep{blei2017variational, jordan1999introduction, ormerod2010explaining} have emerged as a fast and computationally-efficient alternative to MCMC methods for posterior inference in discrete choice models. VB addresses the shortcomings of MCMC by recasting Bayesian inference as an optimisation problem in lieu of a sampling problem. Whilst in MCMC, the posterior distribution of interest is approximated through samples from a Markov chain, VB approximates the posterior distribution of interest through a variational distribution whose parameters are optimised such that the probability distance between the posterior distribution of interest and the approximating variational distribution is minimal. Several studies \citep{bansal2020bayesian, braun2010variational, depraetere2017comparison, tan2017stochastic} derive and assess VB methods for mixed logit models with only inter-individual heterogeneity. All of these studies establish that VB is substantially faster than MCMC at practically no compromises in predictive accuracy. 

Motivated by these recent advances in Bayesian estimation of discrete choice models, this paper has two objectives: First, we derive a VB method for posterior inference in mixed logit models with unobserved inter- and intra-individual heterogeneity. Second, we benchmark the proposed VB method against MSL and MCMC in a simulation study in terms of parameter recovery, predictive accuracy and computational efficiency.

The remainder of this paper is organised as follows. First, we present the mathematical formulation of mixed logit with unobserved inter- and intra-individual heterogeneity (Section \ref{section:model_formulation}). Then, we develop a VB method for this model (Section \ref{section:vb}) and contrast the performance of this method against MSL and MCMC in a simulation study (Section \ref{section:sim_study}). Last, we conclude with a summary and an outline of directions for future research (Section \ref{section:conclusion}). 

\section{Model formulation} \label{section:model_formulation}

The mixed logit model with unobserved inter- and intra-individual heterogeneity \citep[in particular][]{hess2009allowing, hess2011recovery} is established as follows: On choice occasion $t \in \{1, \ldots T \}$, a decision-maker $n \in \{1, \ldots N \}$ derives utility 
\begin{equation} \label{eq:utility}
U_{ntj} = V(\boldsymbol{X}_{ntj}, \boldsymbol{\beta}_{nt}) + \epsilon_{ntj}
\end{equation}
from alternative $j$ in the set $C_{nt}$. Here, $V()$ denotes the representative utility, $\boldsymbol{X}_{ntj}$ is a row-vector of covariates, $\boldsymbol{\beta}_{nt}$ is a collection of taste parameters, and $\epsilon_{ntj}$ is a stochastic disturbance. The assumption $\epsilon_{ntj} \sim \text{Gumbel}(0,1)$ leads to the multinomial logit (MNL) model such that the probability that decision-maker $n$ chooses alternative $j \in C_{nt}$ on choice occasion $t$ can be expressed as 
\begin{equation} \label{eq:mnl_prb}
P(y_{ntj} = 1 \vert \boldsymbol{X}_{ntj}, \boldsymbol{\beta}_{nt},) = \frac{\exp \left \{ V (\boldsymbol{X}_{ntj}, \boldsymbol{\beta}_{nt}) \right \}}{\sum_{k \in C_{nt}}\exp \left \{ V (\boldsymbol{X}_{ntk}, \boldsymbol{\beta}_{nt}) \right \}},
\end{equation} 
where $y_{ntj}$ equals $1$ if alternative $j \in C_{nt}$ is chosen and zero otherwise. 

In equation \ref{eq:utility}, the taste parameters $\boldsymbol{\beta}_{nt}$ are specified as observation-specific. To allow for dependence between repeated observations for the same individual and to accommodate inter-individual taste heterogeneity, it has become standard practice to adopt Revelt's and Train's (\citeyear{revelt1998mixed}) panel estimator for the mixed logit model. Under this specification, taste homogeneity across replications is assumed such that $\boldsymbol{\beta}_{nt} = \boldsymbol{\beta}_{n}$ $\forall t \in \{ 1, \ldots, T \}$. To accommodate intra-individual taste heterogeneity in addition to inter-individual taste heterogeneity, the taste vector $\boldsymbol{\beta}_{nt}$ can be defined as a normal perturbation around an individual-specific parameter $\boldsymbol{\mu}_{n}$, i.e. $\boldsymbol{\beta}_{nt} \sim \mbox{N}(\boldsymbol{\mu}_{n}, \boldsymbol{\Sigma}_{W})$ for $t = 1, \ldots, T$, where $\boldsymbol{\Sigma}_{W}$ is a covariance matrix. The distribution of individual-specific parameters $\boldsymbol{\mu}_{1:N}$ is then also assumed to be multivariate normal, i.e. $\boldsymbol{\mu}_{n} \sim \text{N}(\boldsymbol{\zeta}, \boldsymbol{\Sigma}_{B})$ for $n = 1, \dots, N$, where $\boldsymbol{\zeta}$ is a mean vector and $\boldsymbol{\Sigma}_{B}$ is a covariance matrix. 

Under a fully Bayesian approach, the parameters $\boldsymbol{\zeta}$, $\boldsymbol{\Sigma}_{B}$, $\boldsymbol{\Sigma}_{W}$ are considered to be random, unknown quantities and are thus given priors. Here, we use a normal prior for mean vector $\boldsymbol{\zeta}$ and a marginally-noninformative Huang's half-t prior \citep{akinc2018Bayesian, huang2013Simple} for the covariance matrices $\boldsymbol{\Sigma}_{B}$ and $\boldsymbol{\Sigma}_{W}$.

Stated succinctly, the generative process of mixed logit with unobserved inter- and intra-individual heterogeneity is as follows:
\begin{align}
a_{B,k} \vert A_{B,k} & \sim \text{Gamma}\left( \frac{1}{2}, \frac{1}{A_{B,k}^{2}} \right), k = 1,\dots,K, \label{eq:mixed logit_inter_intra_gen1}\\
a_{W,k} \vert A_{W,k} & \sim \text{Gamma}\left( \frac{1}{2}, \frac{1}{A_{W,k}^{2}} \right), k = 1,\dots,K, \\
\boldsymbol{\Sigma}_{B} \vert \nu_{B}, \boldsymbol{a}_{B} & \sim \text{IW}\left(\nu_{B} + K - 1, 2 \nu_{B} \text{diag}(\boldsymbol{a}_{B}) \right),  \quad \boldsymbol{a}_{B} = \begin{bmatrix} a_{B,1} & \dots & a_{B,K} \end{bmatrix}^{\top}\\
\boldsymbol{\Sigma}_{W} \vert \nu_{W}, \boldsymbol{a}_{W} & \sim \text{IW}\left(\nu_{W} + K - 1, 2 \nu_{W} \text{diag}(\boldsymbol{a}_{W}) \right),  \quad \boldsymbol{a}_{W} = \begin{bmatrix} a_{W,1} & \dots & a_{W,K} \end{bmatrix}^{\top}\\
\boldsymbol{\zeta} \vert \boldsymbol{\xi}_{0},\boldsymbol{\Xi}_{0} & \sim \text{N}(\boldsymbol{\xi}_{0},\boldsymbol{\Xi}_{0}) \\
\boldsymbol{\mu}_{n} \vert \boldsymbol{\zeta}, \boldsymbol{\Sigma}_{B} & \sim \text{N}(\boldsymbol{\zeta}, \boldsymbol{\Sigma}_{B}), n = 1,\dots,N, \\
\boldsymbol{\beta}_{nt} \vert \boldsymbol{\mu}_{n}, \boldsymbol{\Sigma}_{W} & \sim \text{N}(\boldsymbol{\mu}_{n}, \boldsymbol{\Sigma}_{W}), n = 1,\dots,N, t = 1,\dots,T, \\
 y_{nt} \vert \boldsymbol{\beta}_{nt}, \boldsymbol{X}_{nt}  & \sim \text{MNL}(\boldsymbol{\beta}_{nt}, \boldsymbol{X}_{nt}), n = 1,\dots,N,  \ t = 1,\dots,T, \label{eq:mixed logit_inter_intra_gen2}
\end{align} 
where $\{ \boldsymbol{\xi}_{0},\boldsymbol{\Xi}_{0}, \nu_{B}, \nu_{W}, A_{B,1:K}, A_{W,1:K} \}$ are known hyper-parameters, and $\boldsymbol{\theta} = \{\boldsymbol{a}_{B}, \boldsymbol{a}_{W},\boldsymbol{\Sigma}_{B}, \boldsymbol{\Sigma}_{W}, \boldsymbol{\zeta}, \allowbreak  \boldsymbol{\mu}_{1:N}, \allowbreak \boldsymbol{\beta}_{1:N,1:T_n}\}$ is a collection of model parameters whose posterior distribution we wish to estimate. 

The generative process (expressions \ref{eq:mixed logit_inter_intra_gen1}--\ref{eq:mixed logit_inter_intra_gen2}) implies the following joint distribution of the data and the model parameters:
\begin{equation}
\begin{split}
P (\boldsymbol{y}_{1:N}, \boldsymbol{\theta}) = 
& \left ( \prod_{n=1}^{N} \prod_{t=1}^{T_n} 
P(y_{nt} \vert \boldsymbol{\beta}_{nt}, \boldsymbol{X}_{nt}) P(\boldsymbol{\beta}_{nt} \vert \boldsymbol{\mu}_{n}, \boldsymbol{\Sigma}_{W}) \right )
\left ( \prod_{n=1}^{N}  P(\boldsymbol{\mu}_{n} \vert \boldsymbol{\zeta}, \boldsymbol{\Sigma}_{B}) \right ) \\
& 
P(\boldsymbol{\zeta} \vert \boldsymbol{\xi}_{0},\boldsymbol{\Xi}_{0}) 
P(\boldsymbol{\Sigma}_{B} \vert \omega_{B}, \boldsymbol{B}_{B}) 
\left (\prod_{k=1}^{K} P(a_{B,k} \vert s, r_{B,k}) \right) \\
&
P(\boldsymbol{\Sigma}_{W} \vert \omega_{W}, \boldsymbol{B}_{W}) 
\left (\prod_{k=1}^{K} P(a_{W,k} \vert s,r_{W,k}) \right)  
\end{split}
\end{equation}
where 
$\omega_{B} = \nu_{B} + K - 1$, 
$\boldsymbol{B}_{B} = 2\nu_{B} \text{diag}(\boldsymbol{a}_{B})$, 
$\omega_{W} = \nu_{W} + K - 1$, 
$\boldsymbol{B}_{W} = 2\nu_{W} \text{diag}(\boldsymbol{a}_{W})$, 
$s = \frac{1}{2}$,
$r_{B,k} = A_{B,k}^{-2}$ and
$r_{W,k} = A_{W,k}^{-2}$.
By Bayes' rule, the posterior distribution of interest is given by
\begin{equation} \label{chapter_vb_inter_intra:eq:post}
P(\boldsymbol{\theta} \vert \boldsymbol{y}_{1:N}) 
= \frac{P (\boldsymbol{y}_{1:N}, \boldsymbol{\theta})}{\int P (\boldsymbol{y}_{1:N}, \boldsymbol{\theta}) d \boldsymbol{\theta}}
\propto P (\boldsymbol{y}_{1:N}, \boldsymbol{\theta}).
\end{equation}

Exact inference of this posterior distribution is not possible, because the model evidence $\int P (\boldsymbol{y}_{1:N}, \boldsymbol{\theta}) \allowbreak d \boldsymbol{\theta}$ is not tractable. Hence, we resort to approximate inference methods. \citet{becker2018bayesian} propose a Markov chain Monte Carlo (MCMC method for posterior inference in the described model. This method is presented in Appendix \ref{appendix:mcmc}. In the subsequent section, we derive a variational Bayes (VB) method for scalable inference. Mixed logit with unobserved inter- and intra-individual heterogeneity can also be estimated in a frequentist way using maximum simulated likelihood (MSL) estimation \citep[in particular][]{hess2009allowing, hess2011recovery}. The MSL method is described in Appendix \ref{appendix:msl}.

\section{Variational Bayes} \label{section:vb}

This section is divided into two subsections: In Section \ref{subsection:vb_background}, we outline foundational principles of posterior inference with variational Bayes (VB), building on related work by \citet{bansal2020bayesian}.\footnote{We also direct the reader to the statistics and machine learning literature \citep{blei2017variational, ormerod2010explaining, zhang2018advances} for more technical reviews of VB estimation.} In Section \ref{subsection:vb_mixed logit_inter_intra}, we derive a VB method for posterior inference in mixed logit with unobserved inter- and intra-individual heterogeneity.

\subsection{Background} \label{subsection:vb_background}

The general idea underlying VB estimation is to recast posterior inference as an optimisation problem. VB approximates an intractable posterior distribution $P(\boldsymbol{\theta} \vert \boldsymbol{y}) = \frac{P(\boldsymbol{\theta} , \boldsymbol{y})}{\int P(\boldsymbol{\theta} , \boldsymbol{y}) d \boldsymbol{\theta}}$ through a parametric variational distribution $q(\boldsymbol{\theta} \vert \boldsymbol{\nu})$. The optimisation problem consists of manipulating the parameters $\boldsymbol{\nu}$ of the variational distribution such that the probability distance between the posterior distribution of interest and the approximating variational distribution is minimal. In contradistinction, MCMC treats posterior inference as a sampling problem, i.e. the posterior distribution of interest is approximated through samples from a Markov chain whose stationary distribution is the posterior distribution of interest.

Translating posterior inference into an optimisation problem effectively addresses the shortcomings of MCMC \citep[see][]{bansal2020bayesian}. First, only the current estimates of the variational parameters rather than thousands of posterior draws from the Markov chains need to be stored. Second, convergence can be easily assessed by considering the change of the variational lower bound or the estimates of the variational parameters between successive iterations. Last, serial correlation becomes irrelevant, because no samples are taken. 

The Kullback-Leibler (KL) divergence \citep{kullback1951information} provides a measure of the probability distance between the variational distribution $q(\boldsymbol{\theta})$ and the posterior distribution $P(\boldsymbol{\theta} \vert \boldsymbol{y})$. It is defined as 
\begin{equation} 
\begin{split} \label{eq:KL}
D_{\text{KL}} \left (q(\boldsymbol{\theta}) \vert \vert P(\boldsymbol{\theta} \vert \boldsymbol{y}) \right ) 
& = \int \ln \left ( \frac{q(\boldsymbol{\theta})}{P(\boldsymbol{\theta} \vert \boldsymbol{y})} \right ) q(\boldsymbol{\theta}) d q(\boldsymbol{\theta}) \\
& = \mathbb{E}_{q} \left \{ \ln q(\boldsymbol{\theta}) \right \} - \mathbb{E}_{q} \left \{ \ln P(\boldsymbol{\theta} \vert \boldsymbol{y}) \right \} \\
& = \mathbb{E}_{q} \left \{ \ln q(\boldsymbol{\theta}) \right \} - \mathbb{E}_{q} \left \{ \ln P(\boldsymbol{y}, \boldsymbol{\theta}) \right \} + \ln P(\boldsymbol{y}).
\end{split}
\end{equation}
VB seeks to minimise this divergence. Yet, $D_{\text{KL}} \left (q(\boldsymbol{\theta}) \vert \vert P(\boldsymbol{\theta} \vert \boldsymbol{y}) \right )$ is not analytically tractable, because the term $\ln P(\boldsymbol{y})$ lacks a closed-form expression. For this reason, we define an alternative variational lower bound, which is referred to as the evidence lower bound (ELBO):
\begin{equation} \label{eq:ELBO}
\text{ELBO}(q) 
= \mathbb{E}_{q} \left \{ \ln P(\boldsymbol{y}, \boldsymbol{\theta}) \right \}- \mathbb{E}_{q} \left \{ \ln q(\boldsymbol{\theta}) \right \}.
\end{equation}
The ELBO is $\ln P(\boldsymbol{y})$ (which does not depend on $\boldsymbol{\theta}$) minus the KL divergence. Minimising the KL divergence between the variational distribution and the posterior distribution of interest (expression \ref{eq:KL}) is equivalent to maximising the ELBO (expression \ref{eq:ELBO}). Thus, the goal of VB can be re-formulated as 
\begin{equation} 
q^{*}(\boldsymbol{\theta}) = \operatorname*{arg\,max}_{q} \left \{ \text{ELBO}(q) \right \}.
\end{equation}

The functional form of the variational distribution $q(\boldsymbol{\theta})$ needs to be configured by the analyst. The complexity of the variational distribution influences both the expressiveness of the variational distribution (and thus the quality of the approximation to the posterior) as well as the difficulty of the optimisation problem. A computationally-convenient family of variational distributions is the mean-field family of distributions \citep[e.g.][]{jordan1999introduction}. The mean-field assumption imposes independence between blocks of model parameters such that the variational distribution exhibits the form $q(\boldsymbol{\theta}_{1:B}) = \prod_{b=1}^{B} q(\boldsymbol{\theta}_{b})$, where $b \in \{1, \ldots, B\}$ indexes the independent blocks of model parameters. Under the mean-field assumption, the optimal density of each variational factor is $q^{*}(\boldsymbol{\theta}_{b}) \propto \exp \mathbb{E}_{- \boldsymbol{\theta}_{b}} \left \{ \ln P(\boldsymbol{y}, \boldsymbol{\theta}) \right \} $, i.e. the optimal density of each variational factor is proportional to the exponentiated expectation of the logarithm of the joint distribution of $\boldsymbol{y}$ and $\boldsymbol{\theta}$, whereby expectation is calculated with respect to all parameters other than $\boldsymbol{\theta}_{b}$ \citep{blei2017variational, ormerod2010explaining}. If the model of interest has a conditionally-conjugate structure, the optimal densities of all variational factors are known probability distributions. In this case, the variational objective can be maximised with the help of a simple iterative coordinate ascent algorithm \citep{bishop2006pattern}, which involves updating the variational factors one at a time conditionally on the current estimates of the other variational factors. 

\subsection{Variational Bayes for mixed logit with unobserved inter- and intra-individual heterogeneity} \label{subsection:vb_mixed logit_inter_intra}

To apply VB to posterior inference in mixed logit with unobserved inter- and intra-individual heterogeneity, we re-formulate the generative process of the model (see expressions \ref{eq:mixed logit_inter_intra_gen1}--\ref{eq:mixed logit_inter_intra_gen2}) such that the hierarchical dependence between the individual- and the observation-specific taste parameters $\boldsymbol{\mu}_{n}$ and $\boldsymbol{\beta}_{nt}$ is removed. We let
\begin{align}
\boldsymbol{\mu}_{n} \vert \boldsymbol{\zeta}, \boldsymbol{\Sigma}_{B} & \sim \text{N}(\boldsymbol{\zeta}, \boldsymbol{\Sigma}_{B}), n = 1,\dots,N, \\
\boldsymbol{\gamma}_{nt} \vert \boldsymbol{0}, \boldsymbol{\Sigma}_{W} & \sim \text{N}(\boldsymbol{0}, \boldsymbol{\Sigma}_{W}), n = 1,\dots,N, t = 1,\dots,T, 
\end{align}
and define
\begin{equation}
\boldsymbol{\beta}_{nt} = \boldsymbol{\mu}_{n} + \boldsymbol{\gamma}_{nt}, n = 1,\dots,N, t = 1,\dots,T. 
\end{equation}
This modification does not change the model but alters the updates in the VB procedure. If the hierarchical dependence between the individual- and the observation-specific taste parameters is preserved, we find that the non-informative structure of $q^{*}(\boldsymbol{\mu}_{n})$ results in a severe underestimation of the inter-individual covariance $\boldsymbol{\Sigma}_{B}$.\footnote{If $\boldsymbol{\mu}_{n}$ and $\boldsymbol{\beta}_{nt}$ are treated as hierarchically dependent, the optimal density of the conjugate variational factors pertaining to the individual-specific parameters $\boldsymbol{\mu}_{n}$ is $q^{*}(\boldsymbol{\mu}_{n}) \propto \text{Normal}(\boldsymbol{\mu}_{\boldsymbol{\mu}_{n}}, \boldsymbol{\Sigma}_{\boldsymbol{\mu}_{n}})$ with $\boldsymbol{\Sigma}_{\boldsymbol{\mu}_{n}} = \left ( \mathbb{E}_{- \boldsymbol{\mu}_{n}} \left \{ \boldsymbol{\Sigma}_{B}^{-1} \right \} + T \mathbb{E}_{- \boldsymbol{\mu}_{n}} \left \{ \boldsymbol{\Sigma}_{W}^{-1} \right \} \right )^{-1}$ and $\boldsymbol{\mu}_{\boldsymbol{\mu}_{n}} = \boldsymbol{\Sigma}_{\boldsymbol{\mu}_{n}} \Big ( \mathbb{E}_{- \boldsymbol{\mu}_{n}} \big \{ \boldsymbol{\Sigma}_{B}^{-1} \big \} \mathbb{E}_{- \boldsymbol{\mu}_{n}} \{\boldsymbol{\zeta} \} + \allowbreak \mathbb{E}_{- \boldsymbol{\mu}_{n}} \big \{  \boldsymbol{\Sigma}_{W}^{-1} \big \} \allowbreak \sum_{t = 1}^{T} \mathbb{E}_{- \boldsymbol{\mu}_{n}}  \boldsymbol{\beta}_{nt}  \Big )$, where $\boldsymbol{\Sigma}_{\boldsymbol{\mu}_{n}}$ only varies across individuals, if the number of choice occasions differs across individuals.}

The modified posterior distribution is
\begin{equation}
P(\boldsymbol{\theta} \vert \boldsymbol{y}_{1:N}) 
= \frac{P (\boldsymbol{y}_{1:N}, \boldsymbol{\theta})}{\int P (\boldsymbol{y}_{1:N}, \boldsymbol{\theta}) d \boldsymbol{\theta}}
\propto P (\boldsymbol{y}_{1:N}, \boldsymbol{\theta}).
\end{equation} 
with $\boldsymbol{\theta} = \{\boldsymbol{a}_{B}, \boldsymbol{a}_{W},\boldsymbol{\Sigma}_{B}, \boldsymbol{\Sigma}_{W}, \boldsymbol{\zeta}, \boldsymbol{\mu}_{1:N}, \allowbreak \boldsymbol{\gamma}_{1:N,1:T_n}\}$ and 
\begin{equation}
\begin{split}
P (\boldsymbol{y}_{1:N}, \boldsymbol{\theta}) = 
& \left ( \prod_{n=1}^{N} \prod_{t=1}^{T_n} 
P(y_{nt} \vert \boldsymbol{\beta}_{nt}, \boldsymbol{X}_{nt}) P(\boldsymbol{\gamma}_{nt} \vert \boldsymbol{0}, \boldsymbol{\Sigma}_{W}) \right )
\left ( \prod_{n=1}^{N}  P(\boldsymbol{\mu}_{n} \vert \boldsymbol{\zeta}, \boldsymbol{\Sigma}_{B}) \right ) \\
& 
P(\boldsymbol{\zeta} \vert \boldsymbol{\xi}_{0},\boldsymbol{\Xi}_{0}) 
P(\boldsymbol{\Sigma}_{B} \vert \omega_{B}, \boldsymbol{B}_{B}) 
\left (\prod_{k=1}^{K} P(a_{B,k} \vert s, r_{B,k}) \right) \\
&
P(\boldsymbol{\Sigma}_{W} \vert \omega_{W}, \boldsymbol{B}_{W}) 
\left (\prod_{k=1}^{K} P(a_{W,k} \vert s,r_{W,k}) \right)  
\end{split}
\end{equation}
We wish to approximate this posterior distribution through a fitted variational distribution. To this end, we posit a variational distribution from the mean-field family such that
\begin{equation} 
q(\boldsymbol{\theta}) =
\left ( \prod_{n=1}^{N}\prod_{t=1}^{T_n} q(\boldsymbol{\gamma}_{n,t}) \right)
\left ( \prod_{n=1}^N q(\boldsymbol{\mu}_{n}) \right )
q(\boldsymbol{\zeta})
q(\boldsymbol{\Sigma}_{B})
\left ( \prod_{k=1}^{K} q(a_{B,k}) \right )
q(\boldsymbol{\Sigma}_{W})
\left ( \prod_{k=1}^{K} q(a_{W,k}) \right ).
\end{equation}
Under the mean-field assumption, the optimal densities of the variational factors are given by 
$ q^{*}(\boldsymbol{\theta}_{j}) \propto \exp \mathbb{E}_{- \boldsymbol{\theta}_{i}} \left \{ \ln P(\boldsymbol{y}, \boldsymbol{\theta}) \right \} $.
We find that 
$q^{*}(a_{B,k} \vert c_{B}, d_{B,k})$, 
$q^{*}(a_{W,k} \vert c_{W}, d_{W,k})$, 
$q^{*}(\boldsymbol{\Sigma}_{B} \lvert w_{B}, \boldsymbol{\Theta}_{B})$, 
$q^{*}(\boldsymbol{\Sigma}_{W} \lvert w_{W}, \boldsymbol{\Theta}_{W})$ and
$q^{*}(\boldsymbol{\zeta} \lvert \boldsymbol{\mu}_{\boldsymbol_{\zeta}}, \allowbreak \boldsymbol{\Sigma}_{\boldsymbol{\zeta}})$
are common probability distributions (see Appendix \ref{appendix:opt_q}).
However, $q^{*}(\boldsymbol{\mu}_{n})$ and $q^{*}(\boldsymbol{\gamma}_{nt})$ are not members of recognisable families of distributions, because the multinomial logit model lacks a general conjugate prior. For computational convenience, we assume
$q(\boldsymbol{\mu}_{n}) = \text{Normal}(\boldsymbol{\mu}_{\boldsymbol{\mu}_{n}}, \boldsymbol{\Sigma}_{\boldsymbol{\mu}_{n}})$ and
$q(\boldsymbol{\gamma}_{nt}) = \text{Normal}(\boldsymbol{\mu}_{\boldsymbol{\gamma}_{nt}}, \boldsymbol{\Sigma}_{\boldsymbol{\gamma}_{nt}})$.

The evidence lower bound (ELBO) of mixed logit with unobserved inter- and intra-individual heterogeneity is presented in Appendix \ref{appendix:elbo}. 
Because of the mean-field assumption, the ELBO can maximised via an iterative coordinate ascent algorithm. Iterative updates of $q(a_{B,k})$, $q(a_{W,k})$, $(\boldsymbol{\Sigma}_{B})$, $(\boldsymbol{\Sigma}_{W})$ and $(\boldsymbol{\zeta})$ are performed by equating each variational factor to its respective optimal distribution $q^{*}(a_{B,k})$, $q^{*}(a_{W,k})$, $q^{*}(\boldsymbol{\Sigma}_{B})$, $q^{*}(\boldsymbol{\Sigma}_{W})$, $q^{*}(\boldsymbol{\zeta})$. 

Yet, updates of $q(\boldsymbol{\mu}_{n})$ and $q(\boldsymbol{\gamma}_{nt})$ demand special treatment, because there is no closed-form expression for the expectation of the log-sum of exponentials (E-LSE) term in equation \ref{chapter_vb_inter_intra:EofJoint}. The intractable E-LSE term is given by
\begin{equation}
\mathbb{E}_{q} \left \{ g_{nt}(\boldsymbol{\mu}_{n}, \boldsymbol{\gamma}_{nt}) \right \}
\end{equation}
with 
\begin{equation}
g_{nt}(\boldsymbol{\mu}_{n}, \boldsymbol{\gamma}_{nt}) = 
\ln \sum_{j' \in C_{nt}} \exp \left ( \boldsymbol{X}_{ntj'}(\boldsymbol{\mu}_{n} + \boldsymbol{\gamma}_{nt}) \right ).
\end{equation}
\citet{bansal2020bayesian} analyse several methods to approximate the E-LSE term and to update the variational factors corresponding to utility parameters in the context of mixed logit with unobserved inter-individual heterogeneity. In this paper, we use quasi-Monte Carlo (QMC) integration \citep[e.g.][]{bhat2001quasi, dick2010digital, sivakumar2005simulation, train2009discrete} to approximate the E-LSE term and then use quasi-Newton (QN) methods \citep[e.g.][]{nocedal2006numerical} to update $q(\boldsymbol{\mu}_{n})$ and $q(\boldsymbol{\gamma}_{nt})$, as we find that the alternative methods discussed in \citet{bansal2020bayesian} perform poorly for mixed logit with unobserved inter- and intra-individual heterogeneity. 

With QMC integration, the intractable E-LSE terms are approximated by simulation. We have
\begin{equation} \label{chapter_vb_inter_intra:eq:else}
\mathbb{E}_{q} \left \{ g_{nt}(\boldsymbol{\mu}_{n}, \boldsymbol{\gamma}_{nt}) \right \} \approx
\frac{1}{D} \sum_{d = 1}^{D}\ln \sum_{j' \in C_{nt}} \exp \left ( \boldsymbol{X}_{ntj'}(\boldsymbol{\mu}_{n} + \boldsymbol{\gamma}_{nt}) \right ),
\end{equation}
where 
$\boldsymbol{\mu}_{n,d} = \boldsymbol{\mu}_{\boldsymbol{\mu}_{n}} + \mbox{chol}(\boldsymbol{\Sigma}_{\boldsymbol{\mu}_{n}}) \boldsymbol{\xi}^{(\boldsymbol{\mu})}_{n,d}$
and 
$\boldsymbol{\gamma}_{nt,d} = \boldsymbol{\mu}_{\boldsymbol{\gamma}_{nt}} + \mbox{chol}(\boldsymbol{\Sigma}_{\boldsymbol{\gamma}_{nt}}) \boldsymbol{\xi}^{(\boldsymbol{\gamma})}_{nt,d}$.
Moreover, $\boldsymbol{\xi}^{(\boldsymbol{\mu})}_{n,d}$ and $\boldsymbol{\xi}^{(\boldsymbol{\gamma})}_{nt,d}$ denote standard normal simulation draws for the inter- and the intra-individual taste parameters, respectively. Note that the simulation approximation given in expression \ref{chapter_vb_inter_intra:eq:else} is much simpler than the simulation approximation required for MSL estimation (see Appendix \ref{appendix:msl}), because the intra-individual draws need not be conditioned on the inter-individual draws, which is an immediate consequence of the mean-field assumption.

Updates of $q(\boldsymbol{\mu}_{n})$ and $q(\boldsymbol{\gamma}_{nt})$ can then be performed with the help of quasi-Newton methods. We have
\begin{equation} \label{chapter_vb_inter_intra:eq:update_q_mu}
\begin{split}
\left \{ \boldsymbol{\mu}^{*}_{\boldsymbol{\mu}_{n}}, \boldsymbol{\Sigma}^{*}_{\boldsymbol{\mu}_{n}} \right \} =
\operatorname*{arg\,max}_{\boldsymbol{\mu}_{\boldsymbol{\mu}_{n}}, \boldsymbol{\Sigma}_{\boldsymbol{\mu}_{n}}} 
\Bigg \{
& \sum_{t=1}^{T} \left ( \boldsymbol{X}_{nt,y_{nt}}(\boldsymbol{\mu}_{\boldsymbol{\mu}_{n}} + \boldsymbol{\mu}_{\boldsymbol{\gamma}_{nt}}) - \mathbb{E}_{q} \left \{ g_{nt}(\boldsymbol{\mu}_{n}, \boldsymbol{\gamma}_{nt}) \right \} \right ) \\
& -\frac{w_{B}}{2} \mbox{tr} \left (\boldsymbol{\Theta}_{B}^{-1} \boldsymbol{\Sigma}_{\boldsymbol{\mu}_{n}} \right ) - \frac{w_{B}}{2} \boldsymbol{\mu}_{\boldsymbol{\mu}_{n}}^{\top} \boldsymbol{\Theta}_{B}^{-1} \boldsymbol{\mu}_{\boldsymbol{\mu}_{n}} + w_{B} \boldsymbol{\mu}_{\boldsymbol{\mu}_{n}}^{\top} \boldsymbol{\Theta}_{B}^{-1} \boldsymbol{\mu}_{\boldsymbol{\zeta}} \\
& + \frac{1}{2}  \ln \vert \boldsymbol{\Sigma}_{\boldsymbol{\mu}_{n}} \vert
\Bigg \}
\end{split} 
\end{equation}
and
\begin{equation} \label{chapter_vb_inter_intra:eq:update_q_gamma}
\begin{split}
\left \{ \boldsymbol{\mu}^{*}_{\boldsymbol{\gamma}_{nt}}, \boldsymbol{\Sigma}^{*}_{\boldsymbol{\gamma}_{nt}} \right \} =
\operatorname*{arg\,max}_{\boldsymbol{\mu}_{\boldsymbol{\gamma}_{nt}}, \boldsymbol{\Sigma}_{\boldsymbol{\gamma}_{nt}}} 
\Bigg \{
& \left ( \boldsymbol{X}_{nt,y_{nt}}(\boldsymbol{\mu}_{\boldsymbol{\mu}_{n}} + \boldsymbol{\mu}_{\boldsymbol{\gamma}_{nt}}) - \mathbb{E}_{q} \left \{ g_{nt}(\boldsymbol{\mu}_{n}, \boldsymbol{\gamma}_{nt}) \right \}  \right ) \\
& -\frac{w_{W}}{2} \mbox{tr} \left (\boldsymbol{\Theta}_{W}^{-1} \boldsymbol{\Sigma}_{\boldsymbol{\gamma}_{nt}} \right ) - \frac{w_{W}}{2} \boldsymbol{\mu}_{\boldsymbol{\gamma}_{nt}}^{\top} \boldsymbol{\Theta}_{W}^{-1} \boldsymbol{\mu}_{\boldsymbol{\gamma}_{nt}}
+ \frac{1}{2}  \ln \vert \boldsymbol{\Sigma}_{\boldsymbol{\gamma}_{nt}} \vert
\Bigg \},
\end{split} 
\end{equation}
whereby the intractable E-LSE terms $\mathbb{E}_{q} \left \{ g_{nt}(\boldsymbol{\mu}_{n}, \boldsymbol{\gamma}_{nt}) \right \}$ are replaced by the approximation given in expression \ref{chapter_vb_inter_intra:eq:else}.

The algorithm presented in Figure \ref{figure:algo_vb} succinctly summarises the VB method for posterior inference in mixed logit models with unobserved inter- and intra-individual heterogeneity.

\begin{figure}[H]
\begin{algorithm}[H]
\textbf{Initialisation:} \\
Set hyper-parameters: $\nu_{B}$, $\nu_{W}$, $A_{B,1:K}$, $A_{W,1:K}$, $\boldsymbol{\xi}_{0}$, $\boldsymbol{\Xi}_{0}$; \\
Provide starting values: 
$\boldsymbol{\mu}_{\boldsymbol{\zeta}}$, 
$\boldsymbol{\Sigma}_{\boldsymbol{\zeta}}$, 
$\boldsymbol{\mu}_{\boldsymbol{\mu}_{1:N}}$, 
$\boldsymbol{\Sigma}_{\boldsymbol{\mu}_{1:N}}$, 
$\boldsymbol{\Sigma}_{\boldsymbol{\mu}_{1:N}}$, $d_{B,1:K}$, $\boldsymbol{\Theta}_{B}$,
$\boldsymbol{\mu}_{\boldsymbol{\gamma}_{1:N,1:T}}$, 
$\boldsymbol{\Sigma}_{\boldsymbol{\gamma}_{1:N,1:T}}$, 
$\boldsymbol{\Sigma}_{\boldsymbol{\gamma}_{1:N,1:T}}$, $d_{W,1:K}$, $\boldsymbol{\Theta}_{W}$; \\
Generate standard normal quasi-random sequences: $\boldsymbol{\xi}^{(\boldsymbol{\mu})}_{1:N,1:D}$ and $\boldsymbol{\xi}^{(\boldsymbol{\gamma})}_{1:N,1:T,1:D}$; \\
\textbf{Coordinate ascent:} \\
$c_{B} = \frac{\nu_{B} + K}{2}$,
$c_{W} = \frac{\nu_{W} + K}{2}$,
$w_{B} = \nu_{B} + N + K - 1$,
$w_{W} = \nu_{W} + NT + K - 1$;\\
\While{not converged}{
Update $q(\boldsymbol{\mu}_{n})$ $\forall$ $n$ using \ref{chapter_vb_inter_intra:eq:update_q_mu}; \\
Update $q(\boldsymbol{\gamma}_{nt})$ $\forall$ $n,t$ using \ref{chapter_vb_inter_intra:eq:update_q_gamma}; \\
$\boldsymbol{\Sigma}_{\boldsymbol{\zeta}} = \left ( \boldsymbol{\Xi}_{0}^{-1} + N w_{B} \boldsymbol{\Theta}_{B}^{-1} \right )^{-1}$;\\
$\boldsymbol{\mu}_{\boldsymbol{\zeta}} = \boldsymbol{\Sigma}_{\boldsymbol{\zeta}} \left ( \boldsymbol{\Xi}_{0}^{-1} \boldsymbol{\Xi}_{0} + w_{B} \boldsymbol{\Theta}_{B}^{-1} \sum_{n = 1}^{N} \boldsymbol{\mu}_{\boldsymbol{\mu}_{n}}  \right )$; \\
$\boldsymbol{\Theta}_{B} = 2 \nu_{B} \text{diag} \left ( \frac{c_{B}}{\boldsymbol{d_{B}}} \right ) + N \boldsymbol{\Sigma_{\zeta}} + \sum_{n = 1}^{N} \left ( \boldsymbol{\Sigma}_{\boldsymbol{\mu}_{n}} + (\boldsymbol{\mu}_{\boldsymbol{\mu}_{n}} - \boldsymbol{\mu}_{\boldsymbol{\zeta}}) (\boldsymbol{\mu}_{\boldsymbol{\mu}_{n}} - \boldsymbol{\mu}_{\boldsymbol{\zeta}})^{\top} \right)$; \\
$\boldsymbol{\Theta}_{W} = 2 \nu_{W} \text{diag} \left ( \frac{c_{W}}{\boldsymbol{d_{W}}} \right ) + \sum_{n = 1}^{N} \sum_{t = 1}^{T} \left ( \boldsymbol{\Sigma}_{\boldsymbol{\gamma}_{nt}} + \boldsymbol{\mu}_{\boldsymbol{\gamma}_{nt}} \boldsymbol{\mu}_{\boldsymbol{\gamma}_{nt}}^{\top} \right)$; \\
$d_{B,k} = \frac{1}{A_{B,k}^{2}} + \nu_{B} w_{B} \left ( \boldsymbol{\Theta}_{B} ^{-1} \right )_{kk}$ $\forall k $; \\
$d_{W,k} = \frac{1}{A_{W,k}^{2}} + \nu_{W} w_{W} \left ( \boldsymbol{\Theta}_{W} ^{-1} \right )_{kk}$ $\forall k $; \\
}
\end{algorithm}
\caption{Pseudo-code representation of the variational Bayes method for posterior inference in mixed logit models with unobserved inter- and intra-individual heterogeneity} \label{figure:algo_vb}
\end{figure}

\section{Simulation study} \label{section:sim_study}

\subsection{Data and experimental setup}

For the simulation study, we rely on synthetic choice data, which we generate as follows: 
The choice sets comprise five unlabelled alternatives, which are characterised by four attributes.
Decision-makers are assumed to be utility maximisers and to evaluate the alternatives based on the utility specification 
\begin{equation}
U_{ntj} = \boldsymbol{X}_{ntj} \boldsymbol{\beta}_{nt} + \epsilon_{ntj}. 
\end{equation}
Here, $n \in \{ 1, \ldots, N \}$ indexes decision-makers, $t \in \{1, \ldots, T \}$ indexes choice occasions, and $j \in \{1, \ldots, 5 \}$ indexes alternatives. Furthermore, $\boldsymbol{X}_{ntj}$ is a row-vector of uniformly distributed alternative-specific attributes, and $\epsilon_{ntj}$ is a stochastic disturbance sampled from $\text{Gumbel}(0,1)$. 
We consider two experimental scenarios for the generation of the case-specific taste parameters $\boldsymbol{\beta}_{nt}$. In both scenarios, $\boldsymbol{\beta}_{nt}$ are drawn via the following process:
\begin{align}
\boldsymbol{\mu}_{n} \vert \boldsymbol{\zeta}, \boldsymbol{\Sigma}_{B} & \sim \text{N}(\boldsymbol{\zeta}, \boldsymbol{\Sigma}_{B}), n = 1,\dots,N, \\
\boldsymbol{\beta}_{nt} \vert \boldsymbol{\mu}_{n}, \boldsymbol{\Sigma}_{W} & \sim \text{N}(\boldsymbol{\mu}_{n}, \boldsymbol{\Sigma}_{W}), n = 1,\dots,N, t = 1,\dots,T,
\end{align} 
where
$\boldsymbol{\Sigma}_{B} = \text{diag}(\boldsymbol{\sigma}_{B}) \boldsymbol{\Omega}_{B} \text{diag}(\boldsymbol{\sigma}_{B})$ and
$\boldsymbol{\Sigma}_{W} = \text{diag}(\boldsymbol{\sigma}_{W}) \boldsymbol{\Omega}_{W} \text{diag}(\boldsymbol{\sigma}_{W})$.
Here, $\{ \boldsymbol{\sigma}_{B}, \boldsymbol{\sigma}_{W} \}$ represent standard deviation vectors and $\{ \boldsymbol{\Omega}_{B}, \boldsymbol{\Omega}_{W} \}$ are correlation matrices. 
In each scenario, we vary the degree of correlation across the inter- and intra-individual taste parameters. In scenario 1, the degree of correlation is relatively low, whereas it is relatively high in scenario 2. 
In both scenarios, we let 
$\boldsymbol{\sigma}_{B}^{2} = 2 \cdot \frac{2}{3} \cdot \vert \boldsymbol{\zeta} \vert $ and 
$\boldsymbol{\sigma}_{W}^{2} = 2 \cdot \frac{1}{3} \cdot \vert \boldsymbol{\zeta} \vert$,
i.e. the total variance of each random parameter is twice the absolute value of its mean, whereby two thirds of the total variance are due to inter-individual taste variation, and one third of the total variance is due to intra-individual taste variation.
The assumed values of $\boldsymbol{\zeta}$, $\boldsymbol{\Omega}_{B}$ and $\boldsymbol{\Omega}_{W}$ for each scenario are enumerated in Appendix \ref{appendix:true_parameters}. 
In both scenarios, the alternative-specific attributes $\boldsymbol{X}_{ntj}$ are drawn from $\text{Uniform}(0, 2)$, which implies an error rate of approximately 50\%, i.e. in 50\% of the cases decision-makers deviate from the deterministically-best alternative due to the stochastic utility component.
In each scenario, $N$ takes a value in $\{250, 1000 \}$, and $T$ takes a value in $\{8, 16\}$. For each experimental scenario and for each combination of $N$ and $T$, we consider 30 replications, whereby the data for each replication are generated using a different random seed.   

\subsection{Accuracy assessment}

We evaluate the accuracy of the estimation approaches in terms of their ability to recover parameters in finite sample and in terms of their predictive accuracy.

\subsubsection{Parameter recovery}

To assess how the estimation approaches perform at recovering parameters, we calculate the root mean square error (RMSE) for selected parameters, namely the mean vector $\boldsymbol{\zeta}$ and the unique elements $\{ \boldsymbol{\Sigma}_{B,U}, \boldsymbol{\Sigma}_{W,U} \}$ of the covariance matrices $\{ \boldsymbol{\Sigma}_{B}, \boldsymbol{\Sigma}_{W} \}$.
Given a collection of parameters $\boldsymbol{\theta}$ and its estimate $\hat{\boldsymbol{\theta}}$, RMSE is defined as
\begin{equation}
\text{RMSE}(\boldsymbol{\theta}) = \sqrt{ \frac{1}{J} (\hat{\boldsymbol{\theta}} - \boldsymbol{\theta})^{\top} (\hat{\boldsymbol{\theta}} - \boldsymbol{\theta})},
\end{equation}
where $J$ denotes the total number of scalar parameters collected in $\boldsymbol{\theta}$.
For MSL, point estimates of $\boldsymbol{\zeta}$, $\boldsymbol{\Sigma}_{B}$ and $\boldsymbol{\Sigma}_{W}$ are directly obtained. For MCMC, estimates of the parameters of interest are given by the means of the respective posterior draws. 
For VB, we have 
$\hat{\boldsymbol{\zeta}} = \boldsymbol{\mu}_{\boldsymbol{\zeta}}$, 
$\hat{\boldsymbol{\Sigma}}_{B} = \frac{\boldsymbol{\Theta}_{B}}{w_{B} - K - 1}$, and
$\hat{\boldsymbol{\Sigma}}_{W} = \frac{\boldsymbol{\Theta}_{W}}{w_{W} - K - 1}$.
As our aim is to evaluate how well the estimation methods perform at recovering the distributions of the realised individual- and observation-specific parameters $\{ \boldsymbol{\mu}_{1:N}, \boldsymbol{\beta}_{1:N,1:T} \}$, we use the sample mean $\boldsymbol{\zeta}_{0} = \frac{1}{N} \sum_{n = 1}^{N} \boldsymbol{\mu}_{n}$ and the sample covariances $\boldsymbol{\Sigma}_{B,0} = \frac{1}{N} \sum_{n = 1}^{N} (\boldsymbol{\mu}_{n} - \boldsymbol{\zeta}_{0}) (\boldsymbol{\mu}_{n} - \boldsymbol{\zeta}_{0})^{\top}$ and $\boldsymbol{\Sigma}_{W,0} = \frac{1}{NT} \sum_{n = 1}^{N} \sum_{t = 1}^{T} (\boldsymbol{\beta}_{nt} - \boldsymbol{\mu}_{n}) (\boldsymbol{\beta}_{nt} - \boldsymbol{\mu}_{n})^{\top}$ as true parameter values for $\boldsymbol{\zeta}$, $\boldsymbol{\Sigma}_{B}$ and $\boldsymbol{\Sigma}_{W}$, respectively.

\subsubsection{Predictive accuracy}

To assess the predictive accuracy of the Bayesian methods, we consider two out-of-sample prediction scenarios. In the first scenario, we predict choice probabilities for a new set of individuals, i.e. we predict \emph{between} individuals. In the second scenario, we predict choice probabilities for new choice sets for individuals who are already in the sample, i.e. we predict \emph{within} individuals. For each of these scenarios, we calculate the total variation distance \citep[TVD;][]{braun2010variational} between the true and the estimated predictive choice distributions. 

We highlight three important advantages of using TVD as a measure of predictive accuracy. First, TVD is a strictly proper scoring rule \citep{gneiting2007strictly} in that TVD for a given choice set is exclusively minimised by the true predictive choice distribution as opposed to alternative measures of predictive accuracy such as the hit rate. Second, TVD allows for a robust assessment of predictive accuracy, as the estimated predictive choice distribution is compared against the true predictive choice distribution rather than against a single realised choice. Third, TVD fully accounts for the posterior uncertainty captured by the estimation methods. Disregarding posterior uncertainty leads to tighter predictions but also gives a false sense of precision, in particular when the posterior variance of relevant model parameters is large.

We proceed as follows to calculate TVD in the two prediction scenarios:
\begin{enumerate}
\item To evaluate the \emph{between-individual predictive accuracy}, we compute TVD for a validation sample, which we generate together with each training sample. Each validation sample is based on the same data generating process as its respective training sample, whereby the number of decision-makers is set to 25 and the number of observations per decision-maker is set to one. The true predictive choice distribution for a choice set $C_{nt}$ with attributes $\boldsymbol{X}_{nt}^{*}$ from the validation sample is given by
\begin{equation}
P_{\text{true}}(y_{nt}^{*} \vert \boldsymbol{X}_{nt}^{*}) = 
\int \left ( \int P(y_{nt}^{*} = j \vert \boldsymbol{X}_{nt}^{*}, \boldsymbol{\beta}) f(\boldsymbol{\beta} \vert \boldsymbol{\mu}, \boldsymbol{\Sigma}_{W}) d \boldsymbol{\beta} \right ) f(\boldsymbol{\mu} \vert \boldsymbol{\zeta}, \boldsymbol{\Sigma}_{B}) d \boldsymbol{\mu}.
\end{equation}
The true predictive choice distribution is not tractable and is therefore simulated using 2,000 pseudo-random draws for $\boldsymbol{\mu}$ from $\text{N}(\boldsymbol{\zeta}, \boldsymbol{\Sigma}_{B})$ and 2,000 pseudo-random draws for $\boldsymbol{\beta}$ from $\text{N}(\boldsymbol{\mu}, \boldsymbol{\Sigma}_{W})$.
The corresponding estimated predictive choice distribution is 
\begin{equation}
\begin{split}
\hat{P} (y_{nt}^{*} \vert \boldsymbol{X}_{nt}^{*}, \boldsymbol{y}) = & 
\int \int \int
\left (
\int
\left ( \int P(y_{nt}^{*} \vert \boldsymbol{X}_{nt}^{*}, \boldsymbol{\beta}) f(\boldsymbol{\beta} \vert \boldsymbol{\mu}, \boldsymbol{\Sigma}_{W}) d \boldsymbol{\beta} \right ) f(\boldsymbol{\mu}, \boldsymbol{\Sigma}_{B}) d \boldsymbol{\mu}
\right ) \\
&
P(\boldsymbol{\zeta}, \boldsymbol{\Sigma}_{B}, \boldsymbol{\Sigma}_{W} \vert \boldsymbol{y}) d \boldsymbol{\zeta} d \boldsymbol{\Sigma}_{B} d \boldsymbol{\Sigma}_{W}.
\end{split}
\end{equation}
For MCMC, we replace $P(\boldsymbol{\zeta}, \boldsymbol{\Sigma}_{B}, \boldsymbol{\Sigma}_{W} \vert \boldsymbol{y}) d \boldsymbol{\zeta} d \boldsymbol{\Sigma}_{B} d \boldsymbol{\Sigma}_{W}$ with the empirical distribution of the respective posterior draws. For VB, we replace $P(\boldsymbol{\zeta}, \boldsymbol{\Sigma}_{B}, \boldsymbol{\Sigma}_{W} \vert \boldsymbol{y}) d \boldsymbol{\zeta} d \boldsymbol{\Sigma}_{B} d \boldsymbol{\Sigma}_{W}$ with the fitted variational distribution $q(\boldsymbol{\zeta}) q(\boldsymbol{\Sigma}_{B}) q(\boldsymbol{\Sigma}_{W})$ and take 1,000 pseudo-random draws for $\{ \boldsymbol{\zeta}, \boldsymbol{\Sigma}_{B}, \boldsymbol{\Sigma}_{W} \}$ from $q(\boldsymbol{\zeta}) q(\boldsymbol{\Sigma}_{B}) q(\boldsymbol{\Sigma}_{W})$. For both MCMC and VB, we use 200 pseudo-random draws each for $\boldsymbol{\mu}$ and $\boldsymbol{\beta}$.
$\text{TVD}_{B}$ is given by
\begin{equation}
\text{TVD}_{B} = \frac{1}{2} \sum_{j \in C_{nt}} \left | 
P_{\text{true}}(y_{nt}^{*} = j \vert \boldsymbol{X}_{nt}^{*})  - 
\hat{P} (y_{nt}^{*} = j \vert \boldsymbol{X}_{nt}^{*}, \boldsymbol{y})  \right |.
\end{equation}
For succinctness, we calculate averages across decision-makers and choice sets.

\item To evaluate the \emph{within-individual predictive accuracy}, we compute TVD for another validation sample, which we generate together with each training sample. For 25 individuals from the training sample, we generate one additional choice set. Then, the true predictive choice distribution for a choice set $C_{nt}$ with attributes $\boldsymbol{X}_{nt}^{*}$ from the validation sample is given by
\begin{equation}
P_{\text{true}}(y_{nt}^{\dagger} \vert \boldsymbol{X}_{nt}^{\dagger}) = 
\int P(y_{nt}^{\dagger} = j \vert \boldsymbol{X}_{nt}^{\dagger}, \boldsymbol{\beta}) f(\boldsymbol{\beta} \vert \boldsymbol{\mu}_{n}, \boldsymbol{\Sigma}_{W}) d \boldsymbol{\beta}
\end{equation}
This integration is not tractable and is therefore simulated using 10,000 pseudo-random draws for $\boldsymbol{\beta}$ from $\text{N}(\boldsymbol{\mu}, \boldsymbol{\Sigma}_{W})$.
The corresponding estimated predictive choice distribution is 
\begin{equation}
\hat{P} (y_{nt}^{\dagger} \vert \boldsymbol{X}_{nt}^{\dagger}, \boldsymbol{y}) = 
\int \int
\left ( \int P(y_{nt}^{\dagger} \vert \boldsymbol{X}_{nt}^{\dagger}, \boldsymbol{\beta}) f(\boldsymbol{\beta} \vert \boldsymbol{\mu}_{n}, \boldsymbol{\Sigma}_{W}) d \boldsymbol{\beta} \right ) P(\boldsymbol{\mu}_{n}, \boldsymbol{\Sigma}_{W} \vert \boldsymbol{y}) d \boldsymbol{\mu}_{n} d \boldsymbol{\Sigma}_{W}.
\end{equation}
For MCMC, we replace $P(\boldsymbol{\mu}_{n}, \boldsymbol{\Sigma}_{W} \vert \boldsymbol{y})$ with the empirical distribution of the respective posterior draws. For VB, we replace $P(\boldsymbol{\mu}_{n}, \boldsymbol{\Sigma}_{W} \vert \boldsymbol{y})$ with the fitted variational distribution $q(\boldsymbol{\mu}_{n}) q(\boldsymbol{\Sigma}_{W})$ and take 1,000 pseudo-random draws for $\{ \boldsymbol{\mu}_{n}, \boldsymbol{\Sigma}_{W} \}$ from $q(\boldsymbol{\mu}_{n}) q(\boldsymbol{\Sigma}_{W})$. For both MCMC and VB, we use 10,000 pseudo-random draws for $\boldsymbol{\beta}$.
$\text{TVD}_{W}$ is given by
\begin{equation}
\text{TVD}_{W} = \frac{1}{2} \sum_{j \in C_{nt}} \left | 
P_{\text{true}}(y_{nt}^{\dagger} = j \vert \boldsymbol{X}_{nt}^{\dagger})  - 
\hat{P} (y_{nt}^{\dagger} = j \vert \boldsymbol{X}_{nt}^{\dagger}, \boldsymbol{y})  \right |.
\end{equation}
Again, we calculate averages across decision-makers and choice sets for succinctness.
\end{enumerate}

To quantify the benefits of accommodating intra-individual heterogeneity in addition to inter-individual heterogeneity, we also estimate mixed logit models with only inter-individual heterogeneity via MCMC \citep[see][Section 3]{bansal2020bayesian} and compute $\text{TVD}_{B}$ and $\text{TVD}_{W}$ for these models. 

\subsection{Implementation details}

We implement the MSL, MCMC and VB estimators by writing our own Python code.\footnote{The code is publicly available at \url{https://github.com/RicoKrueger/mixed logit_inter_intra}.}

For MSL and VB, the numerical optimisations are carried out with the help of the limited-memory Broyden-Fletcher-Goldfarb-Shanno (L-BFGS) algorithm \citep{nocedal2006numerical} contained in Python's SciPy library \citep{jones2001open} and analytical gradients are supplied. Details concerning the implementation of MSL including the required gradient expressions are provided in Appendix \ref{appendix:msl}. For MSL, we use 200 inter-individual simulation draws per decision-maker and 200 intra-individual simulation draws per observation. For VB, we use 100 inter-individual and 100 intra-individual simulation draws. For both methods, the simulation draws are generated via the Modified Latin Hypercube sampling (MLHS) approach \citep{hess2006use}. To assure that the covariance matrices maintain positive-definiteness, the optimisations are performed with respect to the Cholesky factors of the covariance matrices. 
For VB, we apply the same stopping criterion as \citet{tan2017stochastic}: We define 
$
\boldsymbol{\vartheta} = \begin{bmatrix} \boldsymbol{\zeta}^{\top} & \text{diag}(\boldsymbol{\Theta})^{\top}_{B} & \boldsymbol{d}^{\top}_{B} & \text{diag}(\boldsymbol{\Theta})^{\top}_{W} & \boldsymbol{d}^{\top}_{W} \end{bmatrix}^{\top}
$
and let $\vartheta_{i}^{(\tau)}$ denote the $i$th element of $\boldsymbol{\vartheta}$ at iteration $\tau$. We terminate the iterative coordinate ascent algorithm, when $\delta^{(\tau)} = \operatorname*{arg\,max}_{i} \frac{\vert \vartheta_{i}^{(\tau + 1)} - \vartheta_{i}^{(\tau)}\vert }{\vert \vartheta_{i}^{(\tau)} \vert} < 0.005$. As $\delta^{(\tau)}$ can fluctuate, $\boldsymbol{\vartheta}^{(\tau)}$ is replaced by its average over the last five iterations. 

For MSL and VB, we also take advantage of Python's parallel processing capacities to improve the computational efficiency of the methods. For MSL, we process the likelihood computations in eight parallel batches, each of which corresponds to 25 inter-individual simulation draws. For VB, the updates corresponding to the individual- and the observation-specific parameters are processed in two or eight parallel batches, each of which comprises the observations of 125 individuals. 

For MCMC, the sampler is executed with two parallel Markov chains and 400,000 iterations for each chain, whereby the initial 200,000 iterations of each chain are discarded for burn-in. After burn-in, only every tenth draw is retained to moderate storage requirements and to facilitate post-simulation computations. For mixed logit with only inter-individual heterogeneity, the MCMC sampler is executed with two parallel Markov chains and 100,000 iterations for each chain, whereby the initial 50,000 iterations of each chain are discarded for burn-in. After burn-in, every fifth draw is kept. 

The simulation experiments are conducted on the Katana high performance computing cluster at the Faculty of Science, UNSW Australia.

\subsection{Results}

In Tables \ref{table:results_S1} and \ref{table:results_S2}, we enumerate the simulation results for scenarios 1 and 2, respectively. In each table, we report the means and standard errors of the considered performance metrics across 30 replications for different combinations of sample sizes $N \in \{250,1000\}$ and choice occasions per decision-maker $T \in \{8, 16\}$.

First, we examine the finite-sample properties of the estimators for mixed logit with unobserved inter- and intra-individual heterogeneity. By and large, MSL, MCMC and VB perform equally well at recovering the mean vector $\boldsymbol{\zeta}$ and the covariance matrix $\boldsymbol{\Sigma}_{B}$. For large samples ($N = 1000$), MSL and MCMC perform slightly better than VB at recovering $\boldsymbol{\Sigma}_{B}$, when the number of choice occasions per decision-maker is small ($T = 8$). However, the differences between the methods diminish, as $T$ increases. Furthermore, VB outperforms MCMC and MSL at recovering $\boldsymbol{\Sigma}_{W}$ in the majority of the considered experimental settings. The differences between VB and the two competing methods in the recovery of $\boldsymbol{\Sigma}_{W}$ are particularly pronounced, when the sample size is small ($N = 250$). Overall, the MSL and MCMC estimators perform equally well at recovering the parameters of interest in the considered experimental scenarios. This observation is largely consistent with \citet{becker2018bayesian}. 

Next, we contrast the between-individuals predictive accuracy of the Bayesian methods. Note that in Tables \ref{table:results_S1} and \ref{table:results_S2}, $\text{TVD}_{B}$ is reported in percent. Overall, the differences in between-individuals predictive accuracy between MCMC and VB for mixed logit with unobserved inter-and intra-individual heterogeneity are negligibly small, as the absolute difference in mean $\text{TVD}_{B}$ does not exceed one permille in any of the considered experimental settings. In small samples ($N = 250$), VB performs slightly better than MCMC, but the differences between the methods diminish, as $T$ increases. By contrast, in large samples ($N = 1000$), MCMC performs marginally better than VB. Furthermore, we observe that in all of the considered experimental settings MCMC for mixed logit with only inter-individual heterogeneity is outperformed by the competing methods in term of between-individuals predictive accuracy. This suggests that for the considered data generating process, under which two thirds of the total variance are due to between-individual heterogeneity and the remaining third of the total variance is due to intra-individual heterogeneity, more accurate between-individuals predictions can be produced, if both inter- and intra-individual heterogeneity are accounted for. 

Interestingly, accounting for intra-individual heterogeneity in addition to inter-individual heterogeneity does not result in more accurate within-individuals predictions in any of the considered experimental scenarios. This is because in a parametric hierarchical model, the estimates of the unit-specific parameters are biased towards the population mean and thus do not accurately reflect the true value of the underlying parameter \citep[see][and the literature referenced therein]{danaf2019online}. From the results presented in Tables \ref{table:results_S1} and \ref{table:results_S2}, it can be seen that this bias persists, even if the number of choice tasks per decision-maker is relatively large ($T = 16$). 

Finally, we compare the computational efficiency of the estimation methods for mixed logit with unobserved inter- and intra-individual heterogeneity. Across all of the considered experimental settings, VB is on average between 4.6 and 15.1 times faster than MCMC and between 2.8 and 17.7 times faster than MSL. Interestingly, the MSL method (which uses analytical gradients in combination with parallel processing) is either substantially faster than MCMC or only marginally slower. In small samples ($N = 250$), MSL estimation outperforms MCMC and is between 1.4 and 2.1 times faster than MCMC. In large samples ($N = 1000$), MSL is on par with MCMC and is between 0.9 and 1.1 times faster than MCMC. By contrast, \citet{becker2018bayesian} observe that MSL estimation is substantially slower than MCMC. However, their implementation of MSL estimation relies on numerical gradients and does not use parallel processing. We also highlight that during MCMC estimation, large files for the storage of the posterior draws of $\boldsymbol{\zeta}$, $\boldsymbol{\Sigma}_{B}$, $\boldsymbol{\Sigma}_{W}$ and $\boldsymbol{\mu}_{n}$ for $n = 1, \ldots N$ are produced. For the large samples ($N = 1000$), the posterior draws of both Markov chains require approximately 1.3 Gigabyte of disk space. This is in spite of the fact that only every tenth posterior draw is retained after burn in and also in spite of the fact that the efficient Hierarchical Data Format \citep[HDF;][]{hdf5} is used for the storage of the posterior draws. Both VB and MSL estimation do not consume any disk space during estimation, as no posterior draws need to be stored. 

\begin{landscape}
\begin{table}[h]
\footnotesize
\input{table_results_S1.tex}
\caption{Results for scenario 1 (low correlation)} \label{table:results_S1}
\end{table}
\end{landscape}

\begin{landscape}
\begin{table}[h]
\footnotesize
\input{table_results_S2.tex}
\caption{Results for scenario 2 (high correlation)} \label{table:results_S2}
\end{table}
\end{landscape}

\section{Conclusions} \label{section:conclusion}

Motivated by recent advances in scalable Bayesian inference for discrete choice models, this paper derives a variational Bayes (VB) method for posterior inference in mixed logit models with unobserved inter- and intra-individual heterogeneity. The proposed VB method uses quasi-Monte Carlo (QMC) integration to approximate the intractable expectations of the log-sum of exponentials term and relies on quasi-Newton methods to update the variational factors of the utility parameters. In a simulation study, we benchmark the performance of the proposed VB method against maximum simulated likelihood (MSL) estimation and Markov chain Monte Carlo (MCMC) in terms of parameter recovery, predictive accuracy and computational efficiency. 

In summary, the simulation study demonstrates that VB can be an attractive alternative to MSL and MCMC for the computationally-efficient estimation of mixed logit models with unobserved inter- and intra-individual heterogeneity. In the majority of the considered experimental settings, VB performs at least as well as MSL and MCMC at parameter recovery and out-of-sample prediction, while being between 2.8 and 17.7 times faster than the competing methods. The simulation study also shows that MSL with analytical gradients and parallelised log-likelihood and gradient computations represents a viable alternative to MCMC in terms of both parameter recovery and computational efficiency. This finding stands in contrast to the study by \citet{becker2018bayesian}. In their study, the authors observe that MSL estimation is substantially slower than MCMC, Yet, their implementation of MSL estimation is not parallelised and does not use analytical gradients. 


There are four main directions in which future research can build on the current paper. One direction for future research is to further improve the accuracy of VB. As highlighted by \citet{bansal2020bayesian}, a promising research direction in this regard is to combine the specific benefits of MCMC and VB by marrying the two approaches in an integrated framework \citep[see e.g.][]{salimans2015markov, wolf2016variational}. Another direction for future work is to contrast the performance of the VB, MCMC and MSL estimators for mixed logit with unobserved inter- and intra-individual heterogeneity with the maximum approximate composite marginal likelihood \citep[MACML;][]{bhat2011maximum} estimator for an equivalent mixed probit model \citep[see][]{bhat2011simulation}. A third avenue for future research is to explore how discrete choice models can be formulated so that they can provide accurate individualised recommendations based on personalised predictions. Our analysis suggests that mixed logit models that account for both unobserved inter- and intra-individual heterogeneity do not provide benefits over simpler mixed logit models that only account for inter-individual heterogeneity in regard to within-individual predictions. In the machine learning literature, filtering approaches are used to generate personalised recommendations in recommender systems \citep{koren2009matrix}. Thus, an intriguing avenue for future research is to incorporate such filtering approaches into discrete choice models \citep[also see][]{donnelly2019counterfactual, zhu2019personalized}. Finally, a fourth avenue for future research is to investigate the behavioural sources of what can be detected as intra-individual taste variation and to incorporate explicit representations of the underlying behavioural processes into discrete choice models \citep[see e.g.][]{balbontin2019better}. 

\section*{Author contribution statement}
RK: conception and design, method derivation and implementation, data preparation and analysis, manuscript writing and editing. \\
PB: conception and design, method derivation, data analysis, manuscript editing. \\
MB: conception and design, manuscript editing, supervision. \\
RAD: conception and design, manuscript editing, supervision. \\
THR: conception and design, manuscript editing, supervision, funding acquisition, project administration, resources. \\

\newpage
\bibliographystyle{apalike}
\bibliography{bibliography.bib}

\newpage
\begin{appendices}

\section{Markov chain Monte Carlo estimation} \label{appendix:mcmc}

Here, we present the blocked Gibbs sampler for posterior inference in mixed logit with unobserved inter- and intra-individual heterogeneity, as proposed by \citet{becker2018bayesian}:
\begin{enumerate}
\item Update $a_{B,k}$ for all $k \in \{1, \ldots, K \}$ by sampling $a_{B,k} \sim \text{Gamma} \left ( \frac{\nu_{B} + K}{2}, \frac{1}{A_{B,k}^{2}} + \nu_{B} \left ( \boldsymbol{\Sigma}_{B}^{-1}  \right )_{kk} \right )$.

\item Update $\boldsymbol{\Sigma}_{B}$ by sampling $\boldsymbol{\Sigma}_{B} \sim \text{IW} \Big ( \nu_{B} + N + K - 1,  2 \nu_{B} \text{diag}(\boldsymbol{a}_{B})+  \sum_{n = 1}^{N} (\boldsymbol{\mu}_{n} - \boldsymbol{\zeta}) (\boldsymbol{\mu}_{n} - \boldsymbol{\zeta})^{\top} \Big )$.

\item Update $a_{W,k}$ for all $k \in \{1, \ldots, K \}$ by sampling $a_{W,k} \sim \text{Gamma} \Big ( \frac{\nu_{W} + K}{2}, \frac{1}{A_{W,k}^{2}} + \nu_{W} \left ( \boldsymbol{\Sigma}_{W}^{-1}  \right )_{kk} \Big )$.

\item Update $\boldsymbol{\Sigma}_{W}$ by sampling $\boldsymbol{\Sigma}_{W} \sim \text{IW} \Big ( \nu_{W} + \sum_{n=1}^N  T + K - 1,  2 \nu_{W} \text{diag}(\boldsymbol{a}_{W})+  \sum_{n = 1}^{N} \sum_{t = 1}^{T} (\boldsymbol{\beta}_{nt} - \boldsymbol{\mu}_{n}) (\boldsymbol{\beta}_{nt} - \boldsymbol{\mu}_{n})^{\top} \Big )$.

\item Update $\boldsymbol{\zeta}$ by sampling  $\boldsymbol{\zeta} \sim \text{N}(\boldsymbol{\mu}_{\boldsymbol{\zeta}}, \boldsymbol{\Sigma}_{\boldsymbol{\zeta}})$, 
where
$\boldsymbol{\Sigma}_{\boldsymbol{\zeta}} = \Big ( \boldsymbol{\Xi}_{0}^{-1} + N  \boldsymbol{\Sigma}_{B}^{-1}  \Big )^{-1}$ 
and 
$\boldsymbol{\mu}_{\boldsymbol{\zeta}} = \boldsymbol{\Sigma}_{\boldsymbol{\zeta}} \Big ( \boldsymbol{\Xi}_{0}^{-1} \boldsymbol{\xi}_{0} + \boldsymbol{\Sigma}_{B}^{-1} \sum_{n = 1}^{N}  \boldsymbol{\mu}_{n} \Big )$.

\item Update $\boldsymbol{\mu}_{n}$ for all $n \in \{1, \dots, N \}$ by sampling $\boldsymbol{\mu}_{n} \sim \text{N}(\boldsymbol{\mu}_{\boldsymbol{\mu}_{n}}, \boldsymbol{\Sigma}_{\boldsymbol{\mu}_{n}})$,
where 
$\boldsymbol{\Sigma}_{\boldsymbol{\mu}_{n}} = \Big ( \boldsymbol{\Sigma}_{B}^{-1} + T \boldsymbol{\Sigma}_{W}^{-1}  \Big )^{-1}$ 
and 
$\boldsymbol{\mu}_{\boldsymbol{\mu}_{n}} = \boldsymbol{\Sigma}_{\boldsymbol{\mu}_{n}} \left ( \boldsymbol{\Sigma}_{B}^{-1} \boldsymbol{\zeta} + \boldsymbol{\Sigma}_{W}^{-1} \sum_{t = 1}^{T}  \boldsymbol{\beta}_{nt} \right )$.

\item Update $\boldsymbol{\beta}_{nt}$ for all $n \in \{1, \dots, N \}$ and $t \in \{1, \dots, T \}$:
\begin{enumerate}
\item Propose $\tilde{\boldsymbol{\beta}}_{nt} = \boldsymbol{\beta}_{nt} + \sqrt{\rho} \text{chol}(\boldsymbol{\Sigma}_{W}) \boldsymbol{\eta}$, where $\boldsymbol{\eta} \sim \text{N}(\boldsymbol{0},\boldsymbol{I}_{K})$.
\item Compute $r = 
\frac{P(y_{nt} \vert \boldsymbol{X}_{nt}, \tilde{\boldsymbol{\beta}}_{nt}) \phi( \tilde{\boldsymbol{\beta}}_{nt} \vert \boldsymbol{\mu}_{n},  \boldsymbol{\Sigma}_{W})}
{P(y_{nt} \vert \boldsymbol{X}_{nt}, \boldsymbol{\beta}_{nt}) \phi( \boldsymbol{\beta}_{nt} \vert \boldsymbol{\mu}_{n},  \boldsymbol{\Sigma}_{W})}$.
\item Draw $u \sim \text{Uniform}(0,1)$. If $r \leq u$, accept the proposal. If  $r > u$, reject the proposal. 
\end{enumerate}
\end{enumerate}

Here, $\rho$ is a step size, which needs to be tuned. We employ the same tuning mechanism as \citet{train2009discrete}: $\rho$ is set to an initial value of 0.1 and after each iteration, $\rho$ is decreased by 0.001, if the average acceptance rate across all decision-makers is less than 0.3; $\rho$ is increased by 0.001, if the average acceptance rate across all decision-makers is more than 0.3. 

\section{Maximum simulated likelihood estimation} \label{appendix:msl}

In maximum simulated likelihood (MSL) estimation, the mean vector $\boldsymbol{\zeta}$ and the covariance matrices $\{ \boldsymbol{\Sigma}_{B}, \boldsymbol{\Sigma}_{W} \}$ are treated as fixed, unknown parameters, whereas the individual- and case-specific parameters $\boldsymbol{\mu}_{n}$ and $\boldsymbol{\beta}_{nt}$ are treated as stochastic nuisance parameters. Point estimates of $\{\boldsymbol{\zeta}, \boldsymbol{\Sigma}_{B}, \boldsymbol{\Sigma}_{W} \}$ are obtained via maximisation of the unconditional log-likelihood, whereby the optimisation is in fact performed with respect to the Cholesky factors $\{ \boldsymbol{L}_{B},\boldsymbol{L}_{W} \}$ of the covariance matrices in order to maintain positive-definiteness of $\{ \boldsymbol{\Sigma}_{B}, \boldsymbol{\Sigma}_{W} \}$. 

To formulate the unconditional log-likelihood, we define $\boldsymbol{\beta}_{nt} = \boldsymbol{\mu}_{n} + \boldsymbol{\gamma}_{nt}$, where $\boldsymbol{\mu}_{n} \sim \text{N}(\boldsymbol{\zeta}, \boldsymbol{\Sigma}_{B})$ is an individual-specific random parameter with density $f(\boldsymbol{\mu}_{n} \vert \boldsymbol{\zeta}, \boldsymbol{\Sigma}_{B})$, and where $\boldsymbol{\gamma}_{nt} \sim \text{N}(\boldsymbol{0}, \boldsymbol{\Sigma}_{W})$ is a case-specific random parameter with density $f(\boldsymbol{\gamma}_{nt} \vert \boldsymbol{\Sigma}_{W})$. We then obtain the unconditional log-likelihood by integrating out the stochastic nuisance parameters:
\begin{equation} \label{eq:msl_ll}
LL(\boldsymbol{\theta}) = 
\sum_{n = 1}^{N}
\ln 
\left (
\int 
\prod_{t = 1}^{T}
\left (
\int
P(y_{nt} \vert \boldsymbol{X}_{nt}, \boldsymbol{\beta}_{nt}) 
f(\boldsymbol{\gamma_{nt}} \vert \boldsymbol{\Sigma}_{W}) d \boldsymbol{\gamma_{nt}}
\right )
f(\boldsymbol{\mu_{n}} \vert \boldsymbol{\zeta}, \boldsymbol{\Sigma}_{B}) d \boldsymbol{\mu_{n}}
\right ),
\end{equation}
where $\boldsymbol{\theta} = \{ \boldsymbol{\zeta}, \boldsymbol{L}_{B},\boldsymbol{L}_{W} \}$. 

Since the integrals in expression \ref{eq:msl_ll} are not analytically tractable, we need to resort to simulation to approximate the log-likelihood. The simulated log-likelihood is given by 
\begin{equation}
SLL(\boldsymbol{\theta}) = 
\sum_{n = 1}^{N}
\ln 
\left (
\frac{1}{D} \sum_{d = 1}^{D}
\prod_{t = 1}^{T}
\left (
\frac{1}{R} \sum_{r = 1}^{R}
P(y_{nt} \vert \boldsymbol{X}_{nt}, \boldsymbol{\beta}_{nt,dr}) 
\right )
\right ),
\end{equation}
where $\boldsymbol{\beta}_{nt,dr} = \boldsymbol{\zeta} + \boldsymbol{L}_{B} \boldsymbol{\xi}_{n,d} + \boldsymbol{L}_{W} \boldsymbol{\xi}_{nt,r}$. $\boldsymbol{\xi}_{n,d}$ and $\boldsymbol{\xi}_{nt,r}$ denote standard normal simulation draws. For each decision-maker, we take $D$ draws for $\boldsymbol{\mu}_{n}$ and for each case, we take $R$ draws for $\boldsymbol{\gamma}_{nt}$.

Point estimates $\hat{\boldsymbol{\theta}}$ are then given by
\begin{equation}
\hat{\boldsymbol{\theta}} = 
\operatorname*{arg\,max}_{\boldsymbol{\theta}} SLL(\boldsymbol{\theta}).
\end{equation}
This optimisation problem can be solved with the help of quasi-Newton methods such as the limited-memory BFGS algorithm \citep[see e.g.][]{nocedal2006numerical}. Quasi-Newton methods rely on the gradient of the objective function to find local optima. In principle, gradients can be approximated numerically. However, the numerical approximation of gradients is computationally expensive, as it involves many function evaluations. Computation times can be drastically reduced when analytical or simulated gradients are supplied to the optimiser. In the case of the mixed logit model with inter- and intra-individual heterogeneity, the two levels of integration impose a substantial computational burden, and thus efficient optimisation routines are critical for moderating estimation times. 

In what follows, we derive expressions for the gradients of the mixed logit model with inter- and intra-individual heterogeneity. To the best of our knowledge, this is the first time these gradients are presented in the literature. 
First, we let $\boldsymbol{\vartheta}_{i}$ denote one of the model parameters collected in $\boldsymbol{\theta}$. We have
\begin{equation}
\frac{\partial}{\partial \boldsymbol{\vartheta}_{i}} 
SLL(\boldsymbol{\Theta}) 
= 
\sum_{n = 1}^{N}
\frac
{
\frac{1}{D} \sum_{d = 1}^{D}
\frac{\partial}{\partial \boldsymbol{\varphi}_{i}}
\prod_{t = 1}^{T}
\left (
\frac{1}{R} \sum_{r = 1}^{R}
P(y_{nt} \vert \boldsymbol{X}_{nt}, \boldsymbol{\beta}_{nt,dr}) 
\right )
}
{
\frac{1}{D} \sum_{d = 1}^{D}
\prod_{t = 1}^{T}
\left (
\frac{1}{R} \sum_{r = 1}^{R}
P(y_{nt} \vert \boldsymbol{X}_{nt}, \boldsymbol{\beta}_{nt,dr}) 
\right ).
}
\end{equation}
To find the derivative in the numerator, we define 
\begin{equation}
\psi_{nt,d}(\boldsymbol{\theta})
=
\frac{1}{R} \sum_{r = 1}^{R}
P(y_{nt} \vert \boldsymbol{X}_{nt}, \boldsymbol{\beta}_{nt,dr}) 
\end{equation}
with 
\begin{equation}
\begin{split}
\psi'_{nt,d}(\boldsymbol{\theta})
= & 
\frac{\partial \psi_{nt,dr}(\boldsymbol{\Theta})}{\partial \boldsymbol{\vartheta}_{i}} \\
= &
\frac{1}{R} \sum_{r = 1}^{R}
\Bigg (
P(y_{nt} \vert \boldsymbol{X}_{nt}, \boldsymbol{\beta}_{nt,dr}) 
\frac{\partial V(\boldsymbol{X}_{ntj}, \boldsymbol{\beta}_{nt,dr})}{\partial \boldsymbol{\vartheta}_{i}} \\
& - 
\sum_{j' \in C_{nt}: j' \neq y_{nt} } 
\bigg (
P(y_{nt} \vert \boldsymbol{X}_{nt}, \boldsymbol{\beta}_{nt,dr}) 
P(j' \vert \boldsymbol{X}_{nt}, \boldsymbol{\beta}_{nt,dr}) 
\frac{\partial V(\boldsymbol{X}_{ntj'}, \boldsymbol{\beta}_{nt,dr})}{\partial \boldsymbol{\vartheta}_{i}}
\bigg )
\Bigg ).
\end{split}
\end{equation}
Note that if the representative utility is specified as linear-in-parameters, i.e. 
\begin{equation}
V(\boldsymbol{X}_{ntj}, \boldsymbol{\beta}_{nt,dr}) = \boldsymbol{X}_{ntj} ( \boldsymbol{\zeta} + \boldsymbol{L}_{B} \boldsymbol{\xi}_{n,d} + \boldsymbol{L}_{W} \boldsymbol{\xi}_{nt,r} ),
\end{equation} 
we have 
\begin{align}
\frac{\partial V(\boldsymbol{X}_{ntj}, \boldsymbol{\beta}_{nt,dr})}{\partial \boldsymbol{\zeta}} & = \boldsymbol{X}_{ntj}^{\top}, \\
\frac{\partial V(\boldsymbol{X}_{ntj}, \boldsymbol{\beta}_{nt,dr})}{\partial \boldsymbol{L}_{B}} & = \boldsymbol{X}_{ntj}^{\top} \boldsymbol{\xi}_{n,d}^{\top}, \\
\frac{\partial V(\boldsymbol{X}_{ntj}, \boldsymbol{\beta}_{nt,dr})}{\partial \boldsymbol{L}_{W}} & = \boldsymbol{X}_{ntj}^{\top} \boldsymbol{\xi}_{nt,r}^{\top}.
\end{align} 
From the product rule of differentiation, it follows that
\begin{equation}
\frac{\partial}{\partial \boldsymbol{\vartheta}_{i}}
\prod_{t = 1}^{T}
\left (
\frac{1}{R} \sum_{r = 1}^{R}
P(y_{nt} \vert \boldsymbol{X}_{nt}, \boldsymbol{\beta}_{nt,dr}) 
\right )
= 
\left (
\prod_{t = 1}^{T} 
\psi_{nt,dr}(\boldsymbol{\theta})
\right )
\left (
\sum_{t = 1}^{T}
\frac
{
\psi'_{nt,d}(\boldsymbol{\theta})
}
{
\psi_{nt,d}(\boldsymbol{\theta})
}
\right ).
\end{equation}

\section{Optimal densities of conjugate variational factors} \label{appendix:opt_q}

\subsection{\texorpdfstring{$q^{*}(a_{B,k})$ and $q^{*}(a_{W,k})$}{q*(a_B,k) and q*(a_W,k)}}

\begin{equation}
\begin{split}
q^{*}(a_{B,k}) 
& \propto \exp \mathbb{E}_{- a_{B,k}} \left \{ \ln P \left (a_{B,k} \vert s,  \frac{1}{A_{B,k}^{2}} \right ) + \ln P(\boldsymbol{\Sigma}_{B} \vert \omega_{B}, \boldsymbol{B}_{B}) \right \} \\
& \propto \exp \mathbb{E}_{- a_{B,k}} \left \{ (s - 1) \ln a_{B,k} - \frac{a_{B,k}}{A_{B,k}^{2}} + \frac{\omega_{B}}{2} \ln \boldsymbol{B}_{B,kk} - \frac{1}{2} \boldsymbol{B}_{B,kk} \left ( \boldsymbol{\Sigma}_{B}^{-1} \right )_{kk} \right \} \\
& \propto \exp \left \{ \left ( \frac{\nu_{B} + K}{2} - 1 \right ) \ln a_{B,k} - \left ( \frac{1}{A_{B,k}^{2}} + \nu_{B} \mathbb{E}_{- a_{B,k}} \left \{ \left ( \boldsymbol{\Sigma}_{B}^{-1} \right )_{kk} \right \} \right ) a_{B,k} \right \} \\
& \propto \text{Gamma}(c_{B}, d_{B,k}),
\end{split}
\end{equation}
where 
$c_{B} = \frac{\nu_{B} + K}{2}$ and 
$d_{B,k} = \frac{1}{A_{B,k}^{2}} + \nu_{B} \mathbb{E}_{- a_{B,k}} \left \{ \left ( \boldsymbol{\Sigma}_{B}^{-1} \right )_{kk} \right \}$. 
Furthermore, we note that $\mathbb{E} a_{B,k} = \frac{c_{B}}{d_{B,k}}$.

$q^{*}(a_{W,k})$ can be derived in the same way. We have $q^{*}(a_{W,k}) \propto \text{Gamma}(c_{W}, d_{W,k})$ with 
$c_{W} = \frac{\nu_{W} + K}{2}$ and 
$d_{W,k} = \frac{1}{A_{W,k}^{2}} + \nu_{W} \mathbb{E}_{- a_{W,k}} \left \{ \left ( \boldsymbol{\Sigma}_{W}^{-1} \right )_{kk} \right \}$. Moreover, $\mathbb{E} a_{W,k} = \frac{c_{W}}{d_{W,k}}$.

\subsection{\texorpdfstring{$q^{*}(\boldsymbol{\Sigma}_{B})$ and $q^{*}(\boldsymbol{\Sigma}_{W})$}{q*(Sigma_B) and q*(Sigma_W)}}

\begin{equation} 
\begin{split}
q^{*}(\boldsymbol{\Sigma}_{B}) 
\propto & \exp \mathbb{E}_{- \boldsymbol{\Sigma}_{B}} \left \{ \ln P(\boldsymbol{\Sigma}_{B} \vert \omega_{B}, \boldsymbol{B}_{B}) + \sum_{n=1}^{N} \ln P(\boldsymbol{\mu}_{n} \vert \boldsymbol{\zeta}, \boldsymbol{\Sigma}_{B}) \right \} \\
\propto & \exp \mathbb{E}_{- \boldsymbol{\Sigma}_{B}} \Bigg \{ 
- \frac{\omega_{B} + K + 1}{2} \ln \vert \boldsymbol{\Sigma}_{B} \vert - \frac{1}{2} \text{tr} \left ( \boldsymbol{B}_{B} \boldsymbol{\Sigma}_{B}^{-1} \right ) \\ &
- \frac{N}{2} \ln \vert \boldsymbol{\Sigma}_{B} \vert - \frac{1}{2} \sum_{n = 1}^{N} (\boldsymbol{\mu}_{n} - \boldsymbol{\zeta})^{\top} \boldsymbol{\Sigma}_{B}^{-1} (\boldsymbol{\mu}_{n} - \boldsymbol{\zeta})
\Bigg \} \\
= & \exp \Bigg \{ 
- \frac{\omega_{B} + N + K + 1}{2} \ln \vert \boldsymbol{\Sigma}_{B} \vert \\ &
 - \frac{1}{2} \text{tr} \left ( \boldsymbol{\Sigma}_{B}^{-1} \mathbb{E}_{- \boldsymbol{\Sigma}_{B}} \left \{ \boldsymbol{B}_{B} + 
 \sum_{n = 1}^{N} (\boldsymbol{\mu}_{n} - \boldsymbol{\zeta}) (\boldsymbol{\mu}_{n} - \boldsymbol{\zeta})^{\top} \right \} \right )
\Bigg \} \\
\propto & \text{IW}(w_{B}, \boldsymbol{\Theta}_{B}),
\end{split}
\end{equation}
where 
$w_{B} = \nu_{B} + N + K - 1$ and 
$\boldsymbol{\Theta}_{B} = 
2 \nu_{B} \text{diag} \left ( \frac{c_{B}}{\boldsymbol{d}_{B}} \right )
+ N \boldsymbol{\Sigma_{\zeta}}
+ \sum_{n = 1}^{N} \left (
\boldsymbol{\Sigma}_{\boldsymbol{\mu}_{n}} + (\boldsymbol{\mu}_{\boldsymbol{\mu}_{n}} - \boldsymbol{\mu}_{\boldsymbol{\zeta}}) (\boldsymbol{\mu}_{\boldsymbol{\mu}_{n}} - \boldsymbol{\mu}_{\boldsymbol{\zeta}})^{\top}
 \right)
$.
We use 
$\mathbb{E}\left(\boldsymbol{\mu}_{n} \boldsymbol{\mu}_{n}^{\top}\right) = \boldsymbol{\mu}_{\boldsymbol{\mu}_{n}} \boldsymbol{\mu}_{\boldsymbol{\mu}_{n}}^{\top} + \boldsymbol{\Sigma}_{\boldsymbol{\mu}_{n}}$ and 
$\mathbb{E}\left(\boldsymbol{\zeta}\boldsymbol{\zeta}^{\top}\right) = \boldsymbol{\mu}_{\boldsymbol{\zeta}} \boldsymbol{\mu}_{\boldsymbol{\zeta}}^{\top} + \boldsymbol{\Sigma}_{\boldsymbol{\zeta}}$. 
Furthermore, we note that 
$\mathbb{E} \{ \boldsymbol{\Sigma}_{B}^{-1} \} = w_{B} \boldsymbol{\Theta}_{B}^{-1}$ 
and
$\mathbb{E} \{ \ln \vert \boldsymbol{\Sigma}_{B} \vert \} = \ln \vert  \boldsymbol{\Theta}_{B} \vert + C$, where $C$ is a constant.

$q^{*}(\boldsymbol{\Sigma}_{W})$ can be derived similarly. We have $q^{*}(\boldsymbol{\Sigma}_{W}) \propto \text{IW}(w_{W}, \boldsymbol{\Theta}_{W})$ 
with 
$w_{W} = \nu_{W} + \sum_{n = 1}^{N} T + K - 1$ and 
$\boldsymbol{\Theta}_{W} = 
2 \nu_{W} \text{diag} \left ( \frac{c_{W}}{\boldsymbol{d}_{W}} \right )
+ \sum_{n = 1}^{N} \sum_{t = 1}^{T} \Big (
\boldsymbol{\Sigma}_{\boldsymbol{\beta}_{nt}} + \boldsymbol{\mu}_{\boldsymbol{\beta}_{nt}} \boldsymbol{\mu}_{\boldsymbol{\beta}_{nt}}^{\top}
 \Big)$. 
Moreover, 
$\mathbb{E} \{ \boldsymbol{\Sigma}_{W}^{-1} \} = w_{W} \boldsymbol{\Theta}_{W}^{-1}$ 
and
$\mathbb{E} \{ \ln \vert \boldsymbol{\Sigma}_{W} \vert \} = \ln \vert  \boldsymbol{\Theta}_{W} \vert + C$, where $C$ is a constant.

\subsection{\texorpdfstring{$q^{*}(\boldsymbol{\zeta})$}{q*(zeta)}}

\begin{equation}
\begin{split}
q^{*}(\boldsymbol{\zeta}) 
\propto & \exp \mathbb{E}_{- \boldsymbol{\zeta}} \left \{ \ln P(\boldsymbol{\zeta} \vert \boldsymbol{\xi}_{0},\boldsymbol{\Xi}_{0}) + \sum_{n=1}^{N} \ln P(\boldsymbol{\mu}_{n} \vert \boldsymbol{\zeta}, \boldsymbol{\Sigma}_{B}) \right \} \\
\propto & \exp \mathbb{E}_{- \boldsymbol{\zeta}} \left \{ 
- \frac{1}{2} \boldsymbol{\zeta}^{\top} \boldsymbol{\Xi}_{0}^{-1} \boldsymbol{\zeta} + \boldsymbol{\zeta}^{\top} \boldsymbol{\Xi}_{0}^{-1} \boldsymbol{\xi}_{0}
- \frac{N}{2} \boldsymbol{\zeta}^{\top} \boldsymbol{\Sigma}_{B}^{-1} \boldsymbol{\zeta} + \sum_{n = 1}^{N} \boldsymbol{\zeta}^{\top} \boldsymbol{\Sigma}_{B}^{-1} \boldsymbol{\mu}_{n}
\right \} \\
\propto & \exp \Bigg \{ 
- \frac{1}{2} \Bigg (
\boldsymbol{\zeta}^{\top} \left ( \boldsymbol{\Xi}_{0}^{-1} + N \mathbb{E}_{- \boldsymbol{\zeta}} \left \{ \boldsymbol{\Sigma}_{B}^{-1} \right \} \right ) \boldsymbol{\zeta} \\ &
- 2 \boldsymbol{\zeta}^{\top} \left ( \boldsymbol{\Xi}_{0}^{-1} \boldsymbol{\xi}_{0} + \mathbb{E}_{- \boldsymbol{\zeta}} \left \{  \boldsymbol{\Sigma}_{B}^{-1} \right \} \sum_{n = 1}^{N} \mathbb{E}_{- \boldsymbol{\zeta}} \boldsymbol{\mu}_{n} \Bigg)
\right )
\Bigg \} \\
\propto & \text{Normal}(\boldsymbol{\mu}_{\boldsymbol{\zeta}}, \boldsymbol{\Sigma}_{\boldsymbol{\zeta}}),
\end{split}
\end{equation}
where 
$\boldsymbol{\Sigma}_{\boldsymbol{\zeta}} = \left ( \boldsymbol{\Xi}_{0}^{-1} + N \mathbb{E}_{- \boldsymbol{\zeta}} \left \{ \boldsymbol{\Sigma}_{B}^{-1} \right \} \right )^{-1}$ and 
$\boldsymbol{\mu}_{\boldsymbol{\zeta}} = \boldsymbol{\Sigma}_{\boldsymbol{\zeta}} \left ( \boldsymbol{\Xi}_{0}^{-1} \boldsymbol{\xi}_{0} + \mathbb{E}_{- \boldsymbol{\zeta}} \left \{  \boldsymbol{\Sigma}_{B}^{-1} \right \} \sum_{n = 1}^{N} \mathbb{E}_{- \boldsymbol{\zeta}}  \boldsymbol{\mu}_{n}  \right )$. 
Furthermore, we note that 
$\mathbb{E} \boldsymbol{\zeta} = \boldsymbol{\mu}_{\boldsymbol{\zeta}}$ and $\mathbb{E} \boldsymbol{\mu}_{n} = \boldsymbol{\mu}_{\boldsymbol{\mu}_{n}}$.

\section{Evidence lower bound} \label{appendix:elbo}

The evidence lower bound (ELBO) of mixed logit with unobserved inter- and intra-individual heterogeneity is given by
\begin{equation}
 \text{ELBO} = \mathbb{E}  \left \{ \ln P  (\boldsymbol{y}_{1:N}, \boldsymbol{\theta}) \right \} - \mathbb{E} \left \{ \ln q(\boldsymbol{\theta}) \right \}.
\end{equation}
To derive ELBO, we first define the logarithm of the joint distribution of the data and the unknown model parameters:
\begin{equation}
\begin{split}
\ln P & (\boldsymbol{y}_{1:N}, \boldsymbol{\theta}) \\
= 
& \sum_{n=1}^{N} \sum_{t=1}^{T} \ln P(\boldsymbol{y}_{nt} \vert \boldsymbol{X}_{nt},  \boldsymbol{\beta}_{nt}) \\
& + \sum_{n=1}^{N} \sum_{t=1}^{T} \ln P(\boldsymbol{\gamma}_{nt} \vert \boldsymbol{0},\boldsymbol{\Sigma}_{W})
+ \sum_{n=1}^{N} \ln P(\boldsymbol{\mu}_{n} \vert \boldsymbol{\zeta},\boldsymbol{\Sigma}_{B}) \\
& + \ln P(\boldsymbol{\zeta} \vert \boldsymbol{\xi}_{0},\boldsymbol{\Xi}_{0})
+ \ln P(\boldsymbol{\Sigma}_{B} \vert \omega_{B}, \boldsymbol{B}_{B})
+ \sum_{k=1}^{K} \ln P(a_{B,k} \vert  s,  r_{B,k}) \\
& + \ln P(\boldsymbol{\Sigma}_{W} \vert \omega_{W}, \boldsymbol{B}_{W})
+ \sum_{k=1}^{K} \ln P(a_{W,k} \vert  s,  r_{W,k}) \\
= 
& \sum_{n=1}^{N} \sum_{t=1}^{T} \left ( \boldsymbol{X}_{nt,y_{nt}}(\boldsymbol{\mu}_{n} + \boldsymbol{\gamma}_{nt}) - \ln \sum_{j' \in C_{nt}} \exp \left ( \boldsymbol{X}_{ntj'}(\boldsymbol{\mu}_{n} + \boldsymbol{\gamma}_{nt}) \right ) \right ) \\
& - \frac{NT}{2} \ln \vert \boldsymbol{\Sigma}_{W} \vert -\frac{1}{2} \sum_{n=1}^{N} \sum_{t=1}^{T} \boldsymbol{\gamma}_{nt}^{\top} \boldsymbol{\Sigma}_{W}^{-1} \boldsymbol{\gamma}_{nt} \\
& - \frac{N}{2} \ln \vert \boldsymbol{\Sigma}_{B} \vert -\frac{1}{2} \sum_{n=1}^{N} (\boldsymbol{\mu}_{n} -  \boldsymbol{\zeta})^{\top} \boldsymbol{\Sigma}_{B}^{-1} (\boldsymbol{\beta}_{n} - \boldsymbol{\zeta}) \\
& -\frac{1}{2} (\boldsymbol{\zeta} - \boldsymbol{\xi}_{0})^{\top} \boldsymbol{\Xi}_{0}^{-1} (\boldsymbol{\zeta} - \boldsymbol{\xi}_{0}) \\
& + \frac{\omega_{B}}{2}\ln \vert \boldsymbol{B}_{B}  \vert- \frac{\omega_{B} + K + 1}{2} \ln \vert \boldsymbol{\Sigma}_{B} \vert - \frac{1}{2} \text{tr} \left ( \boldsymbol{B}_{B} \boldsymbol{\Sigma}_{B}^{-1} \right ) \\
& + \sum_{k=1}^{K} \left[(s - 1) \ln a_{B,k} - r_{B,k} a_{B,k} \right] \\
& + \frac{\omega_{W}}{2}\ln \vert \boldsymbol{B}_{W}  \vert- \frac{\omega_{W} + K + 1}{2} \ln \vert \boldsymbol{\Sigma}_{W} \vert - \frac{1}{2} \text{tr} \left ( \boldsymbol{B}_{W} \boldsymbol{\Sigma}_{W}^{-1} \right ) \\
& + \sum_{k=1}^{K} \left[(s - 1) \ln a_{W,k} - r_{W,k} a_{W,k} \right] \\
\end{split}
\end{equation}
After taking expectations, we obtain
\begin{equation} \label{chapter_vb_inter_intra:EofJoint}
\begin{split}
\mathbb{E} & \left \{ \ln P  (\boldsymbol{y}_{1:N}, \boldsymbol{\theta} \right \} \\
= 
& \sum_{n=1}^{N} \sum_{t=1}^{T} \left ( \boldsymbol{X}_{nt,y_{nt}}(\boldsymbol{\mu}_{\boldsymbol{\mu}_{n}} + \boldsymbol{\mu}_{\boldsymbol{\gamma}_{nt}}) - \mathbb{E}_{q} \left \{ \ln \sum_{j' \in C_{nt}} \exp \left ( \boldsymbol{X}_{ntj'}(\boldsymbol{\mu}_{n} + \boldsymbol{\gamma}_{nt}) \right ) \right \} \right ) \\
& - \frac{NT}{2} \ln \vert \boldsymbol{\Theta}_{W} \vert -\frac{w_{W}}{2}\sum_{n=1}^{N} \sum_{t=1}^{T}  \left[\boldsymbol{\mu}_{\boldsymbol{\gamma}_{nt}}^{\top}\boldsymbol{\Theta}_{W}^{-1} \boldsymbol{\mu}_{\boldsymbol{\gamma}_{nt}} + \text{tr} \left ( \boldsymbol{\Theta}_{W}^{-1} \boldsymbol{\Sigma}_{\boldsymbol{\gamma}_{nt}} \right ) \right] \\
& - \frac{N}{2} \ln \vert \boldsymbol{\Theta}_{B} \vert -\frac{w_{B}}{2}\sum_{n=1}^{N} \left[(\boldsymbol{\mu}_{\boldsymbol{\mu}_{n}} - \boldsymbol{\mu}_{\boldsymbol{\zeta}})^{\top}\boldsymbol{\Theta}_{B}^{-1}(\boldsymbol{\mu}_{\boldsymbol{\mu}_{n}} - \boldsymbol{\mu}_{\boldsymbol{\zeta}}) + \text{tr} \left ( \boldsymbol{\Theta}_{B}^{-1} \boldsymbol{\Sigma}_{\boldsymbol{\mu}_{n}} \right ) + \text{tr} \left ( \boldsymbol{\Theta}_{B}^{-1} \boldsymbol{\Sigma}_{\boldsymbol{\zeta}} \right ) \right] \\
& -\frac{1}{2} (\boldsymbol{\mu}_{\boldsymbol{\zeta}} - \boldsymbol{\xi}_{0})^{\top} \boldsymbol{\Xi}_{0}^{-1} (\boldsymbol{\mu}_{\boldsymbol{\zeta}} - \boldsymbol{\xi}_{0})  -\frac{1}{2} \text{tr} \left ( \boldsymbol{\Xi}_{0}^{-1} \boldsymbol{\Sigma}_{\boldsymbol{\zeta}} \right )\\
& - \frac{\omega_{B}}{2} \sum_{k = 1}^{K} \ln d_{B,k} - \frac{\omega_{B} + K + 1}{2} \ln \vert \boldsymbol{\Theta}_{B} \vert - \nu w_{B} \sum_{k = 1}^{K} \frac{c_{B}}{d_{B,k}} \left ( \boldsymbol{\Theta}_{B}^{-1} \right )_{kk} \\
& + \sum_{k = 1}^{K} \left [ (1-s) \ln d_{B,k} - r_{B,k} \frac{c_{B}}{d_{B,k}} \right ] \\
& - \frac{\omega_{W}}{2} \sum_{k = 1}^{K} \ln d_{W,k} - \frac{\omega_{W} + K + 1}{2} \ln \vert \boldsymbol{\Theta}_{W} \vert - \nu w_{W} \sum_{k = 1}^{K} \frac{c_{W}}{d_{W,k}} \left ( \boldsymbol{\Theta}_{W}^{-1} \right )_{kk} \\
& + \sum_{k = 1}^{K} \left [ (1-s) \ln d_{W,k} - r_{W,k} \frac{c_{W}}{d_{W,k}} \right ] \\
\end{split}
\end{equation}
Finally, the negative entropy of the variational distribution is given by
\begin{equation}
\begin{split}
\mathbb{E} \left \{ \ln q(\boldsymbol{\theta}) \right \} =
&  \sum_{k=1}^{K} \ln d_{B,k} 
+ \sum_{k=1}^{K} \ln d_{W,k} \\
& - \frac{K + 1}{2}  \ln \vert \boldsymbol{\Theta}_{B} \vert 
- \frac{K + 1}{2}  \ln \vert \boldsymbol{\Theta}_{W} \vert \\
& - \frac{1}{2} \ln \vert \boldsymbol{\Sigma_{\zeta}} \vert 
- \frac{1}{2} \sum_{n=1}^{N} \ln \vert \boldsymbol{\Sigma}_{\boldsymbol{\mu}_{n}} \vert
- \frac{1}{2} \sum_{n=1}^{N} \sum_{t=1}^{T} \ln \vert \boldsymbol{\Sigma}_{\boldsymbol{\gamma}_{nt}} \vert.
\end{split}
\end{equation}

\section{True population parameters for the simulation study} \label{appendix:true_parameters}

\begin{align}
\boldsymbol{\zeta} & =
\begin{bmatrix}
-0.5 & 0.5 & -0.5 & 0.5
\end{bmatrix}^{\top}
\\
\boldsymbol{\Omega}_{B} & =
\boldsymbol{I}_{4} + \alpha \cdot
\begin{bmatrix}[r]
0 &  0 & 1 & 0 \\
0 &  0 & 0 & 1 \\
1 &  0 & 0 & 0 \\
0 &  1 & 0 & 0 \\
\end{bmatrix}
\\
\boldsymbol{\Omega}_{W} & =
\boldsymbol{I}_{4} + \alpha \cdot
\begin{bmatrix}[r]
0 &  1 & 0 & 1 \\
1 &  0 & 0 & 0 \\
0 &  0 & 0 & 1 \\
1 &  0 & 1 & 0 \\
\end{bmatrix}
\\
\alpha & =
\begin{cases}
0.3 & \text{for scenarios 1 and 3} 
\\
0.6 & \text{for scenarios 2 and 4} \\
\end{cases} 
\end{align}

\end{appendices}

\end{document}

%% file: table_results_S1.tex
\centering
\ra{1.2}
\begin{tabular}{@{} l 
S[table-format=3.1] S[table-format=2.1]  c 
S[table-format=1.4] S[table-format=1.4]  c 
S[table-format=1.4] S[table-format=1.4]  c 
S[table-format=1.4] S[table-format=1.4]  c 
S[table-format=1.4] S[table-format=1.4]  c  
S[table-format=1.4] S[table-format=1.4]  @{}} 
\toprule
& 
\multicolumn{2}{c}{\textbf{Estimation time [min]}} & &
\multicolumn{2}{c}{\textbf{$\text{RMSE}(\boldsymbol{\zeta})$}} & &
\multicolumn{2}{c}{\textbf{$\text{RMSE}(\boldsymbol{\Sigma}_{B,U})$}} & &
\multicolumn{2}{c}{\textbf{$\text{RMSE}(\boldsymbol{\Sigma}_{W,U})$}} & &
\multicolumn{2}{c}{\textbf{$\text{TVD}_{B}$ [\%]}} & &
\multicolumn{2}{c}{\textbf{$\text{TVD}_{W}$ [10\%]}} \\
\cmidrule{2-3} \cmidrule{5-6} \cmidrule{8-9} \cmidrule{11-12}  \cmidrule{14-15} \cmidrule{17-18} 

& 
\textbf{Mean} & \textbf{Std. err.} & &
\textbf{Mean} & \textbf{Std. err.} & &
\textbf{Mean} & \textbf{Std. err.} & &
\textbf{Mean} & \textbf{Std. err.} & &
\textbf{Mean} & \textbf{Std. err.} & &
\textbf{Mean} & \textbf{Std. err.} \\
\midrule

$N = 250$; $T = 8$ \\
\quad MSL &  24.6 &  1.3 & &0.0603 & 0.0041 & &0.1156 & 0.0050 & &0.2240 & 0.0177 & &   NaN &    NaN & &   NaN &    NaN \\
\quad MCMC &  47.4 &  1.0 & &0.0603 & 0.0042 & &0.1130 & 0.0054 & &0.1948 & 0.0214 & &0.5806 & 0.0296 & &0.3686 & 0.0061 \\
\quad VB &   8.9 &  0.2 & &0.0557 & 0.0037 & &0.1038 & 0.0045 & &0.0657 & 0.0030 & &0.5100 & 0.0264 & &0.3669 & 0.0064 \\
\quad MCMC (inter only) &   6.7 &  0.2 & &   NaN &    NaN & &   NaN &    NaN & &   NaN &    NaN & &0.6087 & 0.0227 & &0.3658 & 0.0068 \\
$N = 250$; $T = 16$ \\
\quad MSL &  45.7 &  2.8 & &0.0434 & 0.0030 & &0.0779 & 0.0038 & &0.1480 & 0.0075 & &   NaN &    NaN & &   NaN &    NaN \\
\quad MCMC &  65.8 &  1.4 & &0.0400 & 0.0030 & &0.0705 & 0.0037 & &0.1295 & 0.0068 & &0.5046 & 0.0267 & &0.2963 & 0.0059 \\
\quad VB &  14.4 &  0.3 & &0.0427 & 0.0034 & &0.0700 & 0.0041 & &0.0793 & 0.0024 & &0.4828 & 0.0298 & &0.2959 & 0.0058 \\
\quad MCMC (inter only) &   7.8 &  0.2 & &   NaN &    NaN & &   NaN &    NaN & &   NaN &    NaN & &0.6023 & 0.0215 & &0.2974 & 0.0059 \\
$N = 1000$; $T = 8$ \\
\quad MSL &  97.0 &  3.5 & &0.0279 & 0.0020 & &0.0562 & 0.0020 & &0.1096 & 0.0050 & &   NaN &    NaN & &   NaN &    NaN \\
\quad MCMC & 105.1 &  1.6 & &0.0271 & 0.0021 & &0.0564 & 0.0022 & &0.0991 & 0.0048 & &0.3310 & 0.0117 & &0.3523 & 0.0076 \\
\quad VB &   9.2 &  0.3 & &0.0256 & 0.0019 & &0.0729 & 0.0029 & &0.0580 & 0.0011 & &0.3168 & 0.0119 & &0.3537 & 0.0079 \\
\quad MCMC (inter only) &  12.3 &  0.3 & &   NaN &    NaN & &   NaN &    NaN & &   NaN &    NaN & &0.4592 & 0.0105 & &0.3516 & 0.0080 \\
$N = 1000$; $T = 16$ \\
\quad MSL & 218.2 & 17.4 & &0.0241 & 0.0018 & &0.0386 & 0.0014 & &0.0746 & 0.0036 & &   NaN &    NaN & &   NaN &    NaN \\
\quad MCMC & 203.8 &  3.7 & &0.0207 & 0.0014 & &0.0359 & 0.0014 & &0.0768 & 0.0039 & &0.2677 & 0.0115 & &0.2892 & 0.0071 \\
\quad VB &  13.9 &  0.4 & &0.0245 & 0.0014 & &0.0381 & 0.0015 & &0.0640 & 0.0011 & &0.2573 & 0.0120 & &0.2890 & 0.0071 \\
\quad MCMC (inter only) &  26.0 &  0.6 & &   NaN &    NaN & &   NaN &    NaN & &   NaN &    NaN & &0.4336 & 0.0086 & &0.2902 & 0.0071 \\

\midrule
\multicolumn{18}{l}{
\begin{minipage}[t]{1.4 \textwidth}
Note: 
$\boldsymbol{\zeta}$ = inter-individual mean,
$\boldsymbol{\Sigma}_{B,U}$ = unique elements of inter-individual covariance matrix; 
$\boldsymbol{\Sigma}_{W,U}$ = unique elements of intra-individual covariance matrix; 
$\text{TVD}_{B}$ = total variation distance between true and predicted choice probabilities for a between-individual validation sample;
$\text{TVD}_{W}$ = total variation distance between true and predicted choice probabilities for a within-individual validation sample.
\end{minipage}} \\
\bottomrule
\end{tabular}

%% file: table_results_S2.tex
\centering
\ra{1.2}
\begin{tabular}{@{} l 
S[table-format=3.1] S[table-format=2.1]  c 
S[table-format=1.4] S[table-format=1.4]  c 
S[table-format=1.4] S[table-format=1.4]  c 
S[table-format=1.4] S[table-format=1.4]  c 
S[table-format=1.4] S[table-format=1.4]  c  
S[table-format=1.4] S[table-format=1.4]  @{}} 
\toprule
& 
\multicolumn{2}{c}{\textbf{Estimation time [min]}} & &
\multicolumn{2}{c}{\textbf{$\text{RMSE}(\boldsymbol{\zeta})$}} & &
\multicolumn{2}{c}{\textbf{$\text{RMSE}(\boldsymbol{\Sigma}_{B,U})$}} & &
\multicolumn{2}{c}{\textbf{$\text{RMSE}(\boldsymbol{\Sigma}_{W,U})$}} & &
\multicolumn{2}{c}{\textbf{$\text{TVD}_{B}$ [\%]}} & &
\multicolumn{2}{c}{\textbf{$\text{TVD}_{W}$ [10\%]}} \\
\cmidrule{2-3} \cmidrule{5-6} \cmidrule{8-9} \cmidrule{11-12}  \cmidrule{14-15} \cmidrule{17-18} 

& 
\textbf{Mean} & \textbf{Std. err.} & &
\textbf{Mean} & \textbf{Std. err.} & &
\textbf{Mean} & \textbf{Std. err.} & &
\textbf{Mean} & \textbf{Std. err.} & &
\textbf{Mean} & \textbf{Std. err.} & &
\textbf{Mean} & \textbf{Std. err.} \\
\midrule

$N = 250$; $T = 8$ \\
\quad MSL &  24.2 & 1.5 & &0.0640 & 0.0048 & &0.1161 & 0.0070 & &0.2197 & 0.0163 & &   NaN &    NaN & &   NaN &    NaN \\
\quad MCMC &  47.2 & 0.7 & &0.0639 & 0.0050 & &0.1144 & 0.0081 & &0.2103 & 0.0186 & &0.5917 & 0.0270 & &0.3453 & 0.0072 \\
\quad VB &   9.1 & 0.3 & &0.0552 & 0.0040 & &0.1000 & 0.0050 & &0.1089 & 0.0021 & &0.5253 & 0.0281 & &0.3444 & 0.0072 \\
\quad MCMC (inter only) &   6.4 & 0.2 & &   NaN &    NaN & &   NaN &    NaN & &   NaN &    NaN & &0.6528 & 0.0230 & &0.3452 & 0.0072 \\
$N = 250$; $T = 16$ \\
\quad MSL &  47.5 & 3.0 & &0.0492 & 0.0027 & &0.0829 & 0.0045 & &0.1529 & 0.0099 & &   NaN &    NaN & &   NaN &    NaN \\
\quad MCMC &  62.3 & 1.2 & &0.0438 & 0.0030 & &0.0811 & 0.0044 & &0.1474 & 0.0084 & &0.5124 & 0.0241 & &0.2925 & 0.0059 \\
\quad VB &  13.3 & 0.4 & &0.0480 & 0.0037 & &0.0731 & 0.0036 & &0.1225 & 0.0026 & &0.4955 & 0.0276 & &0.2940 & 0.0058 \\
\quad MCMC (inter only) &   7.6 & 0.2 & &   NaN &    NaN & &   NaN &    NaN & &   NaN &    NaN & &0.6235 & 0.0232 & &0.2962 & 0.0061 \\
$N = 1000$; $T = 8$ \\
\quad MSL & 102.4 & 5.1 & &0.0338 & 0.0023 & &0.0656 & 0.0033 & &0.1166 & 0.0062 & &   NaN &    NaN & &   NaN &    NaN \\
\quad MCMC & 106.9 & 1.7 & &0.0344 & 0.0024 & &0.0639 & 0.0035 & &0.1164 & 0.0065 & &0.3526 & 0.0125 & &0.3507 & 0.0057 \\
\quad VB &   8.9 & 0.2 & &0.0314 & 0.0020 & &0.0846 & 0.0041 & &0.1009 & 0.0014 & &0.3733 & 0.0120 & &0.3539 & 0.0062 \\
\quad MCMC (inter only) &  12.2 & 0.3 & &   NaN &    NaN & &   NaN &    NaN & &   NaN &    NaN & &0.5161 & 0.0112 & &0.3504 & 0.0057 \\
$N = 1000$; $T = 16$ \\
\quad MSL & 231.1 & 8.8 & &0.0201 & 0.0016 & &0.0407 & 0.0016 & &0.0735 & 0.0034 & &   NaN &    NaN & &   NaN &    NaN \\
\quad MCMC & 197.6 & 2.7 & &0.0200 & 0.0017 & &0.0386 & 0.0015 & &0.0803 & 0.0036 & &0.2762 & 0.0107 & &0.2882 & 0.0067 \\
\quad VB &  13.3 & 0.4 & &0.0298 & 0.0019 & &0.0399 & 0.0014 & &0.1108 & 0.0015 & &0.3212 & 0.0094 & &0.2910 & 0.0066 \\
\quad MCMC (inter only) &  24.9 & 0.5 & &   NaN &    NaN & &   NaN &    NaN & &   NaN &    NaN & &0.4948 & 0.0084 & &0.2898 & 0.0066 \\

\midrule
\multicolumn{18}{l}{
\begin{minipage}[t]{1.2 \textwidth}
Note: See Table \ref{table:results_S1} for an explanation of the table headers.
\end{minipage}} \\
\bottomrule
\end{tabular}